\documentclass[11pt]{article}

\usepackage{geometry}

\usepackage[T1]{fontenc}
\usepackage{graphicx}% Standard package for selecting font encodings
\usepackage[utf8x]{inputenc}  % Accept different input encodings
\usepackage{lineno}  % Line numbers on paragraphs
\usepackage{fancyhdr}  % Extensive control of page headers and footers in LaTeX2ε
\usepackage{enumitem}  % Control layout of itemize, enumerate, description
\usepackage{hyperref}  % Extensive support for hypertext in LaTeX
\usepackage{titling}  % Control over the typesetting of the \maketitle command
\usepackage[super,sort&compress]{natbib}  % Flexible bibliography support
\usepackage{mathtools}  % Mathematical tools to use with amsmath
\usepackage{titlesec}  % Select alternative section titles
\usepackage{lastpage}  % Reference last page for Page N of M type footers

\usepackage[english]{babel}  % Multilingual support for Plain TeX or LaTeX
\usepackage{amsmath}  % AMS mathematical facilities for LaTeX
\usepackage{amsfonts}  % TeX fonts from the American Mathematical Society
\usepackage{amssymb}  % Additional symbols from American Mathematical Society
\usepackage{wasysym}  % LaTeX support file to use the WASY2 fonts
\usepackage{bbm}  % "Blackboard-style" cm fonts
\usepackage{array}  % Extending the array and tabular environments
\usepackage{xr}  % References to other LaTeX documents
\usepackage{verbatim} % Reimplementation of and extensions to LaTeX verbatim
\usepackage{float} % Improved interface for floating objects

\hypersetup{%
  pdfborderstyle={/S/U/W 1}% border style will be underline of width 1pt
}

\titleformat{name=\section}{\normalfont\Large\bfseries}{}{0pt}{}
\titleformat{name=\subsection}{\normalfont\large\bfseries}{}{0pt}{}
\titleformat{name=\subsubsection}{\normalfont\normalsize\bfseries}{}{0pt}{}

\newcommand{\email}[1]{\href{mailto:{#1}}{{#1}}}
\newcommand{\keywords}[1]{\textbf{Keywords}: {#1}}

\newcommand{\optincludegraphics}[2][]{}
\newcommand{\optinput}[1]{}
\newtagform{brackets}{[}{]}
\usetagform{brackets}

\newcommand{\thejournal}[1]{Magnetic Resonance in Medicine}

\title{High Fidelity Deep Learning-based MRI Reconstruction with Instance-wise Discriminative Feature Matching Loss}
\immediate\write18{texcount -sum=1,1,1,1,1,1,1 -merge -incbib -dir -utf8 \jobname.tex > \jobname.wcTotal }
\immediate\write18{texcount -sub=none -merge -incbib -dir -utf8 \jobname.tex | grep _main_ | grep -oE '[0-9]+' | python -c "import sys; print(sum(int(l) * w for l, w in zip(sys.stdin, [1, 1, 0, 0, 0, 0, 0])))" > \jobname.wcManuscript }
\immediate\write18{texcount -sub=none -merge -incbib -dir -utf8 \jobname.tex | grep Abstract | grep -oE '[0-9]+' | python -c "import sys; print(sum(int(l) * w for l, w in zip(sys.stdin, [1, 1, 0, 0, 0, 0, 0])))" > \jobname.wcAbstract }

\newcommand{\wcTotal}{\clearpage{\noindent\large{\bf Detailed Word Count} (not to be included for submission)}\verbatiminput{\jobname.wcTotal}}
\newcommand{\wcManuscript}{\input{\jobname.wcManuscript}}
\newcommand{\wcAbstract}{\input{\jobname.wcAbstract}}

\begin{document}

% ======================================================================

\begin{titlepage}
{\noindent\LARGE\bf \thetitle}

\bigskip

% : Insert author names, affiliations and corresponding author email
% : (do not include titles, positions, or degrees).
%FIXME
\begin{flushleft}\large
	Ke Wang\textsuperscript{1,4,*},
	Jonathan I Tamir\textsuperscript{2},
	Alfredo De Goyeneche\textsuperscript{1},
	Uri Wollner\textsuperscript{3},
	Rafi Brada\textsuperscript{3},
	Stella Yu\textsuperscript{1,4},
	Michael Lustig\textsuperscript{1}
\end{flushleft}

\bigskip

\noindent
%FIXME
\begin{enumerate}[label=\textbf{\arabic*}]
\item {Electrical Engineering and Computer Sciences, University of California at Berkeley, California, USA}
\item {Electrical and Computer Engineering, The University of Texas at Austin, Austin, Texas, USA}
\item{GE Global Research, Herzliya, Israel}
\item {International Computer Science Institute, University of California, Berkeley, Berkeley, California, USA}
\end{enumerate}

\bigskip

% : Use the dagger symbol to denote a single equal contribution authorship.
% : Multiple equal-contribution authorship may be included in the acknowledgments.
%FIXME
% \textbf{{†}}: These authors contributed equally to this work.

% : Use the asterisk to denote corresponding authorship.
% : Provide email address in note below.
%FIXME
\textbf{*} Corresponding author:

\indent\indent
\begin{tabular}{>{\bfseries}rl}
Name		& Ke Wang 
\\
Department	& Electrical Engineering and Computer Sciences													\\
Institute	& University of California at Berkeley														\\
         	& California														\\
			& 94720
			\\
            & United States														\\
E-mail		& \email{kewang@berkeley.edu}											\\
\end{tabular}

\vfill

Approximate word count: 250 (Abstract) 5000 (body)

\end{titlepage}

\maketitle

\begin{abstract}

\noindent\textbf{Purpose:}
To improve reconstruction fidelity of fine structures and textures in deep learning (DL) based reconstructions.

\noindent\textbf{Methods:} 
A novel patch-based Unsupervised Feature Loss (UFLoss) is proposed and incorporated into the training of DL-based reconstruction frameworks in order to preserve perceptual similarity and high-order statistics. The UFLoss provides instance-level discrimination by mapping similar instances to similar low-dimensional feature vectors and is trained without any human annotation. By adding an additional loss function on the low-dimensional feature space during training, the reconstruction frameworks from under-sampled or corrupted data can reproduce more realistic images that are closer to the original with finer textures, sharper edges, and improved overall image quality. The performance of the proposed UFLoss is demonstrated on unrolled networks for accelerated 2D and 3D knee MRI reconstruction with retrospective under-sampling. Quantitative metrics including NRMSE, SSIM, and our proposed UFLoss were used to evaluate the performance of the proposed method and compare it with others.

\noindent\textbf{Results:} In-vivo experiments indicate that adding the UFLoss encourages sharper edges and more faithful contrasts compared to traditional and learning-based methods with pure $\ell_2$ loss. More detailed textures can be seen in both 2D and 3D knee MR images. Quantitative results indicate that reconstruction with UFLoss can provide comparable NRMSE and a higher SSIM while achieving a much lower UFLoss value.

\noindent\textbf{Conclusion:} 
We present UFLoss, a patch-based unsupervised learned feature loss, which allows the training of DL-based reconstruction to obtain more detailed texture, finer features, and sharper edges with higher overall image quality under DL-based reconstruction frameworks. \footnote{Code available at: \url{https://github.com/mikgroup/UFLoss}}
 
\noindent\keywords{image reconstruction, compressed sensing, convolutional neural network (CNN), deep learning, feature loss}
\end{abstract}

\section{Introduction}

Magnetic resonance imaging (MRI) offers tremendous benefits to both science and medicine, but unfortunately, MRI data acquisition is inherently time-consuming. As a result, there is great interest in reconstructing diagnostic quality images from limited measurements to shorten scan times. Over the past decades, numerous computational approaches have been proposed to address this problem, including parallel imaging (PI) \cite{sodickson1997simultaneous,pruessmann1999sense,griswold2002generalized} and compressed sensing (CS) \cite{lustig2007sparse}. PI leverages multiple receiver coils to acquire multiple-view images simultaneously for efficient image reconstruction. CS incorporates prior information about the system and signal to constrain the image reconstruction. Both PI and CS have successfully enabled a broad range of clinical applications, and all major MRI vendors have implemented products based on them.

Nonetheless, there remain several challenges with PI and CS.  {\bf 1)} The regularization functions used in CS are hand-crafted (e.g., sparse transformation) or rely on simple learned features (e.g., dictionary learning \cite{ravishankar2010mr}), which are known to be suboptimal at modeling the underlying data distribution \cite{hammernik2018learning}.  {\bf 2)} CS reconstruction is sensitive to the tuning parameters. {\bf 3)} The reconstruction time of CS is relatively long due to iterative optimization.

To overcome these limitations, end-to-end deep learning (DL)-based reconstruction methods \cite{chen2018variable,mardani2018deep,quan2018compressed,schlemper2017deep,hammernik2018learning,aggarwal2018modl,tamir2019unsupervised} have been proposed to learn the regularization terms directly from a large training dataset. Two representative approaches include the Variational Network (VN) \cite{hammernik2018learning} and Model-based Deep Learning (MoDL) \cite{aggarwal2018modl}. Both methods consist of unrolling a conventional iterative CS reconstruction and replacing the regularization step with learnable activation functions or Convolutional Neural Networks (CNNs). End-to-end training is performed in a supervised learning manner. These unrolled learning-based methods have shown great potential at further accelerating reconstruction from under-sampled k-space measurements, well beyond the capabilities of combined parallel imaging and compressed sensing (PICS).

It is well-known that the performance of DL-based methods is dependent on the loss function used for training. The most commonly used loss functions for training are pixel-wise $\ell_1$, $\ell_2$ and the patch-wise structural similarity index (SSIM) \cite{wang2004image} losses \cite{aggarwal2018modl,hammernik2018learning,chen2018variable}. However, these loss functions are usually hand-crafted or based on local statistics, which do not necessarily capture the perceptual information of fine structures, which results in images with degraded perceptual quality and blurring when compared to un-accelerated scans \cite{mardani2018deep,yi2019generative}. 

To address these issues, Generative Adversarial Networks (GANs) \cite{goodfellow2014generative,Isola_2017_CVPR,mirza2014conditional} with adversarial losses have been proposed to exploit the implicit feature information by incorporating discriminators into the reconstruction pipeline \cite{liu2019santis,mardani2018deep,yi2019generative}. Unfortunately, 
%GAN's adversarial loss function is known to be less stable and difficult to optimize due to the large number of free parameters and hand-picked stopping criterion. 
GANs are notoriously hard to train, easily fall into mode collapse, and are sensitive to hyperparameter selections.
Additionally, the adversarial loss is a less-constrained instance-to-set loss function, where improper training parameters may result in unexpected hallucinations and artifacts in reconstructions \cite{sandino2020compressed}. 

Aside from the adversarial loss, recent works in computer vision have shown that CNN-based perceptual losses can be used to learn high-level image feature representations \cite{johnson2016perceptual,zhang2018unreasonable}.  These perceptual loss functions are based on feature layers of classification networks (such as the VGG Net \cite{simonyan2014very}). They are typically designed to work for natural images with a fixed channel number (RGB) and are usually trained in a supervised manner with human-annotated labels, e.g. from ImageNet \cite{deng2009imagenet}. Therefore, simply using perceptual VGG losses may not be ideal for MRI reconstruction tasks. For MR data sets, the dimensionality of the data can vary from application to application  (e.g., 2D/3D complex-valued data, 2D/3D dynamic data), while at the same time, human-annotated labels for MR images are much harder to obtain. More importantly, it is also unclear what kind of human annotations would be best for comparing the image quality for MR images.

\begin{figure}[!ht]
\begin{center}
\includegraphics[width=14cm]{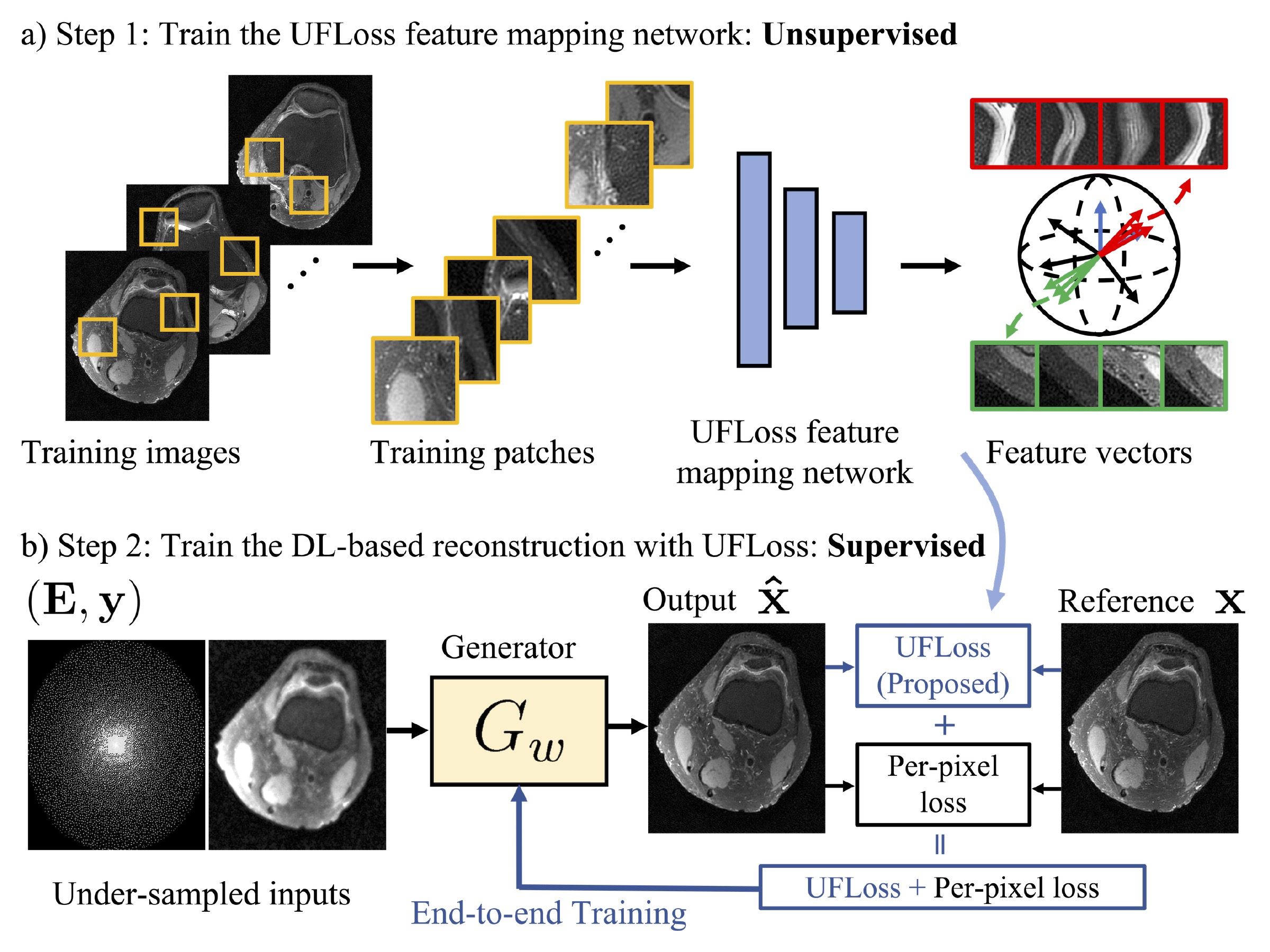}
\end{center}
\caption{Overview of training the DL-based reconstruction with UFLoss. We split the pipeline into two steps. a) Step 1: We pre-train the UFLoss feature mapping network on fully-sampled image patches without human annotations, where the aim of the training is to maximally separate out all the patches in the feature space. b) Step 2: For the training of the DL-based reconstruction, $G_{w,\mathbf{E}}$ represents a reconstruction network with learnable parameters $w$, and given system encoding operator $\mathbf{E}$. The inputs of $G_{w,\mathbf{E}}$ are under-sampled k-space $\mathbf{y}$, and zero-filled reconstruction $\mathbf{E^Hy}$. We feed-forward $\mathbf{E^Hy}$ through $G_{w,\mathbf{E}}$ to obtain the output reconstruction results. We adopt the pre-trained UFLoss network from (a) to compute the UFLoss in the feature space. Then, end-to-end training is performed with respect to the combination of UFLoss and per-pixel loss. Note that the training of DL-based reconstruction with UFLoss is still supervised.}
\label{fig:OV}
\end{figure}

In this work, we propose a novel unsupervised learned feature loss (Figure \ref{fig:OV}) to capture the perceptual and high-order statistical difference within MR images, which we call {\it Unsupervised Feature Loss} (UFLoss). 
The UFLoss is a large-patch-wise loss function that provides instance-level discrimination by mapping similar patches to similar low-dimensional feature vectors using a pre-trained mapping network (which we refer to as UFLoss feature mapping network or UFLoss network) \cite{wanghigh}. 
The rationale of using features from large-patches (typically 40$\times$40 pixels for a 300$\times$300 pixels image) is that we want our UFLoss to capture mid-level structural and semantic features instead of using small patches (typically around 10$\times$10 pixels), which only contain local edge information. On the other hand, we avoid using global features due to the fact that our training set (typically around 5000 slices) is usually not large enough to capture common and general features at a large-image scale. 

Different from adversarial loss, UFLoss is a more-constrained instance-to-instance loss function, which leads to more stable training and reduced hallucinations. Meanwhile, unlike the VGG perceptual loss, pre-training the UFLoss network requires no supervision, and thus is able to capture high-level structural information specifically for MR images without any human annotations.  
Similar to the VGG perceptual loss, UFLoss can also be easily incorporated into the training of DL-based reconstruction networks without modifying the network architecture. Figure~\ref{fig:OV} shows the overall pipeline for using our UFLoss to train a DL-based reconstruction. We first pre-train the UFLoss network on fully sampled image patches without accompanying annotated labels (Figure~\ref{fig:OV}a). This step maps patches to a lower-dimensional space while attempting to maximally separate them in the feature space. The outcome is that {\it similar} patches end up being close together in the feature space while dissimilar ones end up further apart.  This pre-trained feature mapping network is then adopted to compute the UFLoss during the training of the DL-based reconstruction (Figure~\ref{fig:OV}b), which corresponds to the $\ell_2$ distance in the feature space summed across all images patches.  End-to-end training is performed with respect to a combination of UFLoss and per-pixel $\ell_1$/$\ell_2$ or SSIM losses.

To demonstrate the power of UFLoss, we focus on a representative unrolled DL-based reconstruction framework: MoDL \cite{aggarwal2018modl}. 
% The patch-level features of the UFLoss focus on local image structures.  We hypothesize that this locality leads to a more stable training of the DL-based reconstructions with reduced hallucinations compared to other global feature losses, e.g., global adversarial loss.  
We conduct experiments to show that UFLoss is a valid loss function sensitive to increasing low-level intensity deformation. Our results for patch retrieval and patch correlation in MR images demonstrate that \emph{visually similar} patches are indeed close in the feature space.

Our experiments on 2D and 3D in-vivo data show that the addition of the UFLoss encourages more realistic reconstructions with more subtle details and improved overall image quality compared to conventional and learning-based methods with other losses (pure $\ell_2$ loss and $\ell_2$+VGG perceptual loss).

\section{Theory}

\subsection{Unrolled reconstruction for under-sampled MRI}
In conventional under-sampled MRI, the PICS inverse problem can be formulated as \cite{lustig2007sparse}:
\begin{linenomath*}
\begin{equation}
    \hat{\mathbf{x}} = \arg\min_{\mathbf x} \frac{1}{2}\left\lVert \mathbf{Ex}-\mathbf{y}\right\rVert^2_2 +\lambda Q(\mathbf{x}),
    \label{eq:imdl}
\end{equation}
\end{linenomath*}
where $\mathbf{x}$ is the image to be reconstructed, and $\mathbf{y}$ is the measured data in k-space. $\mathbf{E}$ describes the system encoding matrix, which can be further expanded to: $\mathbf{E = UFS}$, where $\mathbf{F}$ is the Fourier transform operator, $\mathbf{S}$ represents the multiple sensitivity maps, and $\mathbf{U}$ corresponds to the k-space sampling operator. For the Cartesian case, \textbf{U} is a diagonal matrix with 1's corresponding to collected k-space and 0's to un-acquired k-space. For non-Cartesian, \textbf{U} is a k-space re-sampling operator from a Cartesian grid to the acquired non-Cartesian trajectory. The goal of this problem is to reconstruct the image which has the lowest error compared to the measured k-space data in the least-squares sense. However, when the sampling rate is below the Nyquist rate, Equation \ref{eq:imdl} becomes ill-posed. Therefore, a regularization term $Q(\mathbf{x})$ with a weighting parameter $\lambda$, which incorporates prior knowledge about the image, is added to constrain the optimization problem. For conventional CS MRI, $Q(\mathbf{x})$ is often chosen to promote sparsity in a certain transform domain such as wavelets or finite spatial differences.

A number of first-order iterative methods have been developed for efficiently solving the minimization problem in  Equation \ref{eq:imdl} for the case where $Q(\mathbf x)$ is convex \cite{beck2009fast,boyd2011distributed}. To further develop fast and high-fidelity reconstructions, recent methods have attempted to directly learn the proximal function $Q$ and the corresponding parameters from a large set of fully-sampled training data in an unrolled fashion \cite{chen2018variable,mardani2018deep,quan2018compressed,schlemper2017deep,hammernik2018learning,aggarwal2018modl,tamir2019unsupervised}.

A widely used unrolled reconstruction framework is MoDL \cite{aggarwal2018modl}, where the reconstruction is formulated as:
\begin{linenomath*}
\begin{equation}
    \hat{\mathbf{x}} = \arg\min_{\mathbf x} \left\lVert \mathbf{Ex}-\mathbf{y}\right\rVert^2_2 +\lambda \left\lVert \mathbf{x}-D_w(\mathbf{x})\right\rVert^2_2.
    % \label{eq:imdl}
\end{equation}
\end{linenomath*}
In this formulation, $D_w$ is a learned CNN denoiser/artifact removal network and $w$ are the learned weighting parameters. The CNN-based prior $\left\lVert \mathbf{x}-D_w(\mathbf{x})\right\rVert^2_2$ results in high values when $\mathbf{x}$ is corrupted by noise and aliasing. Similar to ADMM \cite{boyd2011distributed}, we can solve the optimization problem in the following half-quadratic splitting steps:
\begin{linenomath*}
\begin{equation}
    \mathbf{z}^{k} =D_w(\mathbf{x}^{k})
    \label{eq:5}
\end{equation}
\end{linenomath*}

\begin{linenomath*}
\begin{equation}
\begin{aligned}
\mathbf{x}^{(k+1)} &= \arg\min_{\mathbf x} \left\lVert \mathbf{Ex}-\mathbf{y}\right\rVert^2_2 +\lambda \left\lVert \mathbf{x}-\mathbf{z}^k \right\rVert^2_2\\
&=(\mathbf{E^HE+\lambda\mathbf{I}})^{-1}(\mathbf{E^Hy}+\lambda\mathbf{z}^k)
\end{aligned}
\label{eq:6}
\end{equation}
\end{linenomath*}

Equation \ref{eq:6} can be solved using the Conjugate Gradient (CG) Method while Equation \ref{eq:5} is viewed as a CNN-based forward-pass step. MoDL is formulated as an unrolled network, where in each iteration, a CG layer is followed by a CNN-based proximal step. The unrolled reconstruction can be denoted as $\hat{\mathbf{x}}=G_w(\mathbf{y},\mathbf{E})$, where $\mathbf{y}$, $\mathbf{E}$ and $w$ correspond to the under-sampled k-space measurements, the encoding matrix, and the learnable weights of the reconstruction network, respectively. Training the unrolled model becomes supervised learning with a pre-defined loss function:
\begin{linenomath*}
\begin{equation}
    \min_{w}\sum_{i}\mathcal{L}( G_w(\mathbf{y}_i,\mathbf{E}_i),\mathbf{x}_i),
    \label{eq:loss_plain}
\end{equation}
\end{linenomath*}
where $\mathbf{x}_i$ is the $i^\mathrm{th}$ fully-sampled ground truth image, and $\mathbf{y}_i$ is the retrospectively under-sampled k-space computed by applying the encoding matrix $\mathbf{E}_i$ to generate $\mathbf{y}_i = \mathbf{E}_i\mathbf{x}_i$. The loss function $\mathcal{L}(\cdot)$ can be combinations of $\ell_1$, $\ell_2$, SSIM, and other losses. Once trained, a new under-sampled scan denoted by ${\mathbf{y}}$ with the encoding operator ${\mathbf{E}}$ is reconstructed as:
\begin{linenomath*}
\begin{equation}
    \hat{\mathbf{x}} = G_w({\mathbf{y}},{\mathbf{E}}).
    \label{eq:inference}
\end{equation}
\end{linenomath*}

%%%%%%%%%%%%%%%%%%%%%%%%%%%%%%%%%%%%%%%%%%
\subsection{UFLoss feature mapping network}

\begin{figure}[!ht]
\begin{center}
\includegraphics[width=14cm]{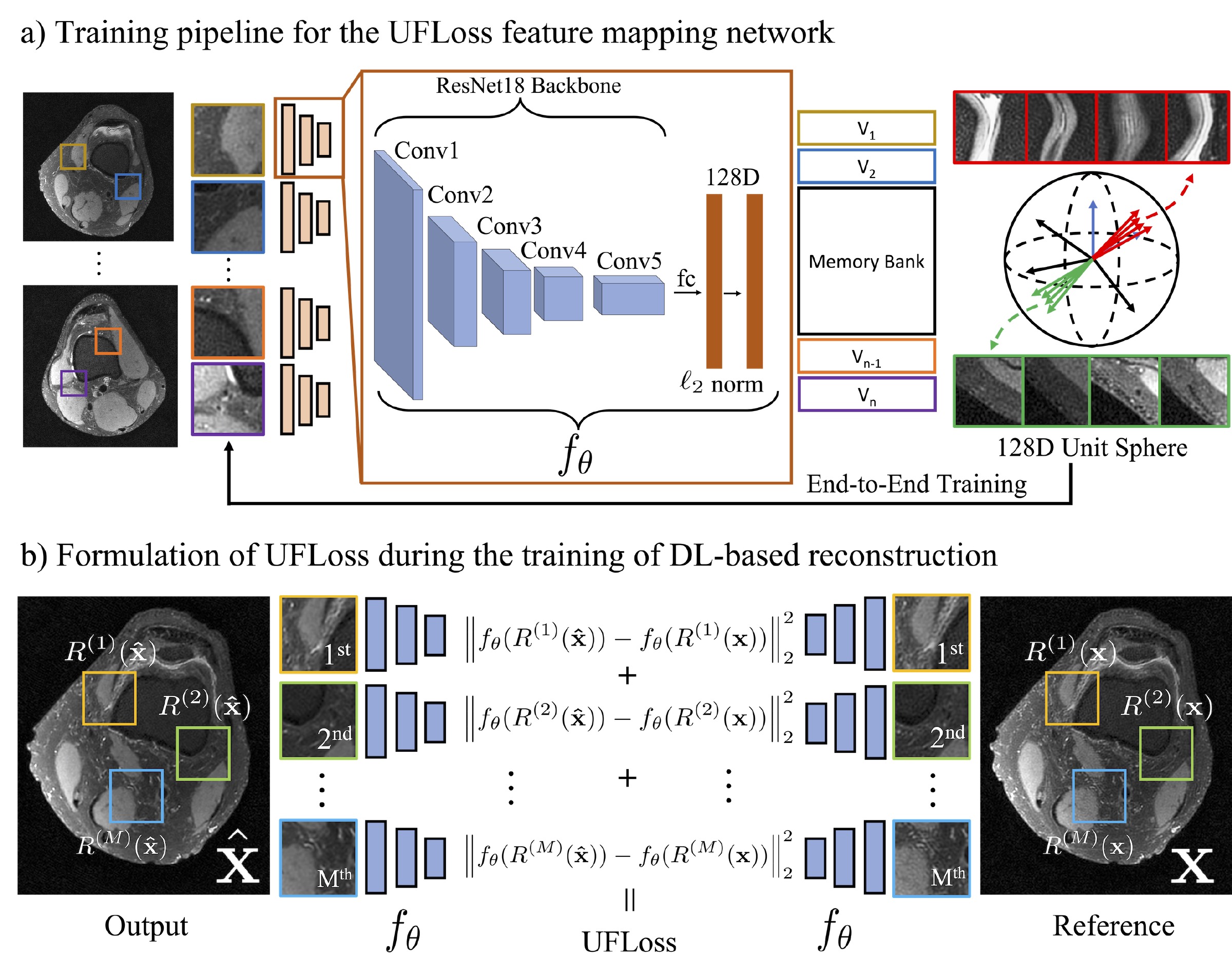}
\end{center}
\caption{a) Training pipeline for the UFloss feature mapping network. Patches cropped from the fully sampled images are separately passed through a ResNet 18 \cite{he2016deep} backbone followed by an $\ell_2$ normalization layer to map the patches to features on a low-dimensional unit sphere (128-dimension unit-norm features in this work). A memory bank is used to store the features from all the training patches to save computation when computing the softmax loss function (Equation \ref{eq:3}). Then, end-to-end training is performed such that each patch is maximally separated from other patches in the 128D unit-norm feature space. Similar patches will naturally cluster in the low-dimensional space. b) Detailed formulation of the proposed UFLoss during the training of the DL-based reconstruction. Operator $R$ extracts a total of $M$ patches from an image. These patches are extracted on a grid with a sliding window. Each patch from the reconstructed output and the fully-sampled reference will go through a pre-trained network $f_\theta$ and mapped to a low-dimensional feature space. The UFLoss corresponds to the sum of the $\ell_2$ distance between the feature vectors from the output and the fully-sampled reference.}
\label{fig:FA}
\end{figure}

As shown in Figure \ref{fig:FA}a), a patch-wise mapping network (UFLoss feature mapping network) is trained to map patches from image-space to a low-dimensional unit-norm feature space, aiming to capture high-level structural differences. The UFLoss network can then be used for training a DL-based reconstruction. In contrast to conventional supervised computer vision tasks, the UFLoss network is trained from fully sampled image patches in an unsupervised fashion. In other words, the training does not use any human annotation, which has been challenging to obtain in large-scale MRI datasets. The training is motivated by contrastive learning \cite{wu2018unsupervised}, where a feature mapping function  $f_\theta$ is learned such that each patch is maximally separated from other patches in a lower-dimensional hypersphere feature space.

Mathematically, we formulate our unsupervised feature mapping using the softmax criterion. Suppose we have $N$ patches $\{\mathbf{p}_1,\mathbf{p}_2,...,\mathbf{p}_N\}$ cropped from the fully sampled images from the training set, with their corresponding unit-norm features $\{\mathbf{v}_1,\mathbf{v}_2,...,\mathbf{v}_N\}$ with $\mathbf{v}_i=f_{\theta}(\mathbf{p}_i)\in\mathbb{R}^d$. For a certain patch $\mathbf{p}$ with feature $\mathbf{v}=f_{\theta}(\mathbf{p})$, the probability of it being identified as the $i^\mathrm{th}$ patch under a linear classifier is:
\begin{linenomath*}
\begin{equation}
    P(i|\mathbf{v})=\frac{\exp{(\mathbf{w}_i^\mathbf{T}\mathbf{v}})}{\sum_{j=1}^N\exp{(\mathbf{w}_j^\mathbf{T}\mathbf{v})}},
    \label{eq:1}
\end{equation}
\end{linenomath*}
where $\mathbf{w}_j$ is the weight vector of class $j$ (or patch $j$), and $\mathbf{w}_j^\mathbf{T}\mathbf{v}$ shows how well the feature vector $\mathbf{v}$ matches the $j^\mathrm{th}$ patch. However, the above formulation  Equation \ref{eq:1} requires a class prototype $\mathbf{w}$ in addition to the patch feature itself, making direct comparison between patches infeasible.  To address this problem, we follow the approach in \cite{wu2018unsupervised} to turn the instance-wise classification into a metric learning problem,  where     $\mathbf{w}_j^\mathbf{T}\mathbf{v}$ in  Equation \ref{eq:1} is replaced with $\mathbf{v}_j^\mathbf{T}\mathbf{v}$.  That is, the $j^\mathrm{th}$ patch feature {\it is} its class prototype itself.  The probability then becomes:
\begin{linenomath*}
\begin{equation}
    P(i|\mathbf{v})=\frac{\exp{(\mathbf{v}_i^\mathbf{T}\mathbf{v})/\tau}}{\sum_{j=1}^N\exp{({\mathbf{v}_j^\mathbf{T}\mathbf{v}})/\tau}},
    \label{eq:2}
\end{equation}
\end{linenomath*}
where $\tau$ is a temperature parameter that controls the extent of separation/concentration of the distribution in the feature space. The learning objective is set to maximize the joint probability $\Pi^{N}_{i=1}P_{\theta}(i|f_{\theta}(\mathbf{x}_i))$, which is equivalent to minimizing the negative log-likelihood over the training set:
\begin{linenomath*}
\begin{equation}
    J(\theta) = - \sum^N_{i=1}\log P(i|f_{\theta}(\mathbf{x_i})).
    \label{eq:3}
\end{equation}
\end{linenomath*}
Note that in order to compute the probability $P(i|\mathbf{v})$ in  Equation \ref{eq:2}, features $\{\mathbf{v}_i\}$ from all the patches are required. Instead of exhaustively computing all the features every time, a memory bank $\mathbf{V} = \{\mathbf{v}_1, \dots, \mathbf{v}_N\}$ is constructed to store all the feature vectors.  During each training iteration, while the network parameters $\theta$ are optimized over the $i^\mathrm{th}$ patch, the $i^\mathrm{th}$ entry of the memory bank $\mathbf{v}_i$ is replaced by the output of the feature mapping network $\mathbf{f}_\theta(\mathbf{
p}_i)\rightarrow\mathbf{v}_i$.

Once trained, the UFLoss network can be used as a perceptual loss term in other supervised reconstruction tasks, as described next.

%%%%%%%%%%%%%%%%%%%%%%%%%%%%%%%%%%%%%%%%%%%%%%%%%%%%%
\subsection{Deep learning-based reconstruction with UFLoss}

The UFLoss network is designed to maximally separate patches in the low-dimensional unit-sphere feature space.  Perceptually similar patches are mapped to similar features. 

Consider the under-sampled reconstruction using an unrolled network in  Equation \ref{eq:loss_plain}.  Suppose we have the ground truth fully-sampled image ${\mathbf{x}_i}$, and the output of the unrolled network $\hat{\mathbf{x}}_i = G_w(\mathbf{y}_i,\mathbf{E}_i)$. Since the inputs of the UFLoss network are image patches (Figure \ref{fig:FA}b), we first extract \textit{M} overlapping image patches from both $\mathbf{x}_i$ and $\hat{\mathbf{x}}_i$, obtaining two patch groups: $\{\mathbf{p}_i^1,\mathbf{p}_i^2,...,\mathbf{p}_i^M\}$ and $\{\hat{\mathbf{p}}_i^1,\hat{\mathbf{p}}_i^2,...,\hat{\mathbf{p}}_i^M\}$. The patches are extracted on a grid with ${N}_s$ pixel strides horizontally and vertically.

During each training step, random shifts between 0 to ${N}_s$ pixels are applied with equal shifts to both $\mathbf{x}_i$ and $\hat{\mathbf{x}}_i$. This choice has the effect of averaging out the blocking artifacts and achieves the same performance as extracting all the patches \cite{lustig2007sparse, tamir2017t2}.

Since we use inner products to measure the distance in the hyperspherical feature space, the UFLoss can be formulated as the average of the negative inner products over all the patches. On top of that, we add a constant 1 in front of our loss function:
\begin{linenomath*}
\begin{equation}
    L_{UFLoss}({\mathbf{x}_i},\hat{\mathbf{x}}_i) = \frac{1}{M}\sum_j 1-\langle f_\theta(\mathbf{p}_i^j),f_\theta(\hat{\mathbf{p}}_i^j)\rangle,
\end{equation}
where $\langle\cdot,\cdot\rangle$ is the inner product operation between two unit-norm vectors and $f_{\theta}$ is the pre-trained UFLoss mapping network. As both $f_{\theta}(\mathbf{p}_i^j)$ and $f_\theta(\hat{\mathbf{p}}_i^j)$ have unit norms, the above loss function can be also written as a mean-squared-error (MSE) in the feature space, or:
\end{linenomath*}

\begin{linenomath*}
\begin{equation}
\begin{aligned}
L_{UFLoss}({\mathbf{x}_i},\hat{\mathbf{x}}_i) &= \frac{1}{M}\sum_j 1-\langle f_\theta(\mathbf{p}_i^j),f_\theta(\hat{\mathbf{p}}_i^j)\rangle\\
&=\frac{1}{2M}\sum_j\left\lVert f_\theta(\mathbf{p}_i^j) \right\rVert^2_2-2\langle f_\theta(\mathbf{p}_i^j),f_\theta(\hat{\mathbf{p}}_i^j)\rangle+\left\lVert f_\theta(\hat{\mathbf{p}}_i^j) \right\rVert^2_2\\
&=\frac{1}{2M}\sum_j\left\lVert f_\theta(\mathbf{p}_i^j)-f_\theta(\hat{\mathbf{p}}_i^j) \right\rVert^2_2.
\end{aligned}
\label{eq:ufloss}	
\end{equation}
\end{linenomath*}
Following the per-pixel $\ell_2$ loss and UFLoss mentioned above, the full objective loss function for the DL-based reconstruction can be written as:
\begin{linenomath*}
\begin{equation}
\begin{aligned}
    L_{Recon} &= L_{MSE-all}+2\mu L_{UFLoss-all}\\
    &= \sum_{i}L_{MSE}({\mathbf{x}_i},\hat{\mathbf{x}}_i)+2\mu\sum_{i}L_{UFLoss}({\mathbf{x}_i},\hat{\mathbf{x}}_i)\\
    &=\sum_{i}\left\lVert G_w(\mathbf{y}_i,\mathbf{E}_i)-\mathbf{x}_i\right\rVert^2_2+\mu\sum_{i}\frac{1}{M}\sum_j\left\lVert f_\theta(\mathbf{p}_i^j)-f_\theta(\hat{\mathbf{p}}_i^j) \right\rVert^2_2,
\end{aligned}
\label{eq:L_recon_full}
\end{equation}
\end{linenomath*}
where $\mu$ is the weighting factor on the contribution of the UFLoss.  End-to-end training is then performed on this total loss to optimize the reconstruction network $G_w$.

\section{Methods}
\subsection{Imaging datasets}

% Add more
We trained and evaluated our proposed UFLoss on both 2D and 3D fully-sampled knee datasets with retrospective under-sampling. We used the fastMRI \cite{zbontar2018fastmri} high resolution knee data set for our 2D experiments.
A total of 5700 fully-sampled slices from 380 cases were split into 320 cases (6080 slices) for training, 40 cases (640 slices) for validation, and 20 cases (320 slices) for testing. Image normalization was performed such that the 95\% percentile of the intensity values was scaled to 1 for each subject. The training dataset includes data from two different contrasts: proton-density with (PDFS) and without (PD) fat suppression. Relevant imaging parameters are described in the fastMRI \cite{zbontar2018fastmri} paper. For the unrolled reconstruction task, retrospective under-sampling was performed by applying a 1D five times accelerated random under-sampling mask (20\% sampling rate) with an 8\% fully sampled k-space center. Sensitivity maps were computed using ESPIRiT \cite{uecker2014espirit} using BART \cite{uecker2015berkeley} with a 24$\times24$ calibration region.

We conducted our 3D experiments on 20 fully sampled 3D knee scans (available at mridata.org) \cite{sawyer2013creation} with retrospective under-sampling.  The k-space data was acquired on a 3T GE Discovery MR 750, with an 8-channel HD knee coil. Scan parameters include a matrix size of 320$\times$320$\times$256, and TE/TR of 25ms/1550ms. A total of 5120 slices from 16 cases were used for training, 640 slices from 2 cases were used for validation, and 640 slices from the remaining 2 cases were used for testing. We normalized each 3D volume with respect to the 95\% percentile of the intensity values for the entire volume. Each 3D volume was under-sampled with a different 8$\times$ Poisson-disk sampling mask (12.5\% sampling rate) with a 24$\times$24 calibration region. Sensitivity maps were computed using ESPIRiT \cite{uecker2014espirit} with a 24$\times24$ calibration region using BART \cite{uecker2015berkeley}. Note that we train both the UFLoss network and the DL-based reconstructions on the entire training set.

\subsection{Implementation of UFLoss feature mapping network}
In all our networks, the input complex-valued MR images/patches $\mathbf{x}\in\mathbb{C}^\mathbf{N}$ are converted into a two-channel representation $\mathbf{x}\in\mathbb{R}^\mathbf{2N}$, where the real and imaginary components are treated as two individual channels. As illustrated in Figure \ref{fig:NW}a), we implemented the UFLoss network using a ResNet 18 \cite{he2016deep} backbone followed by a $\ell_2$ normalization layer to map the input patches to 128 dimension unit-norm features. Based on the FOV and resolution difference, the input patch sizes of the 2D fastMRI knee dataset and 3D knee dataset were set to 60$\times$60 and 40$\times$40 pixels, respectively. The UFLoss networks for the 2D fastMRI and 3D knee datasets were trained separately due to the differences in image content. Eighty patches were extracted from each slice at random locations, resulting in 409,600 patches used to train the UFLoss network. Other hyperparameters include  temperature $\tau$ of $1$ (Equation \ref{eq:2}), batch size $16$, the number of epochs of $100$, and the learning rate of $1\mathrm{e}{-4}$ with Adam \cite{kingma2014adam} optimizer.

\subsection{Implementation of deep learning-based reconstruction with UFLoss}

\begin{figure}[!htb]
\begin{center}
\includegraphics[width=14cm]{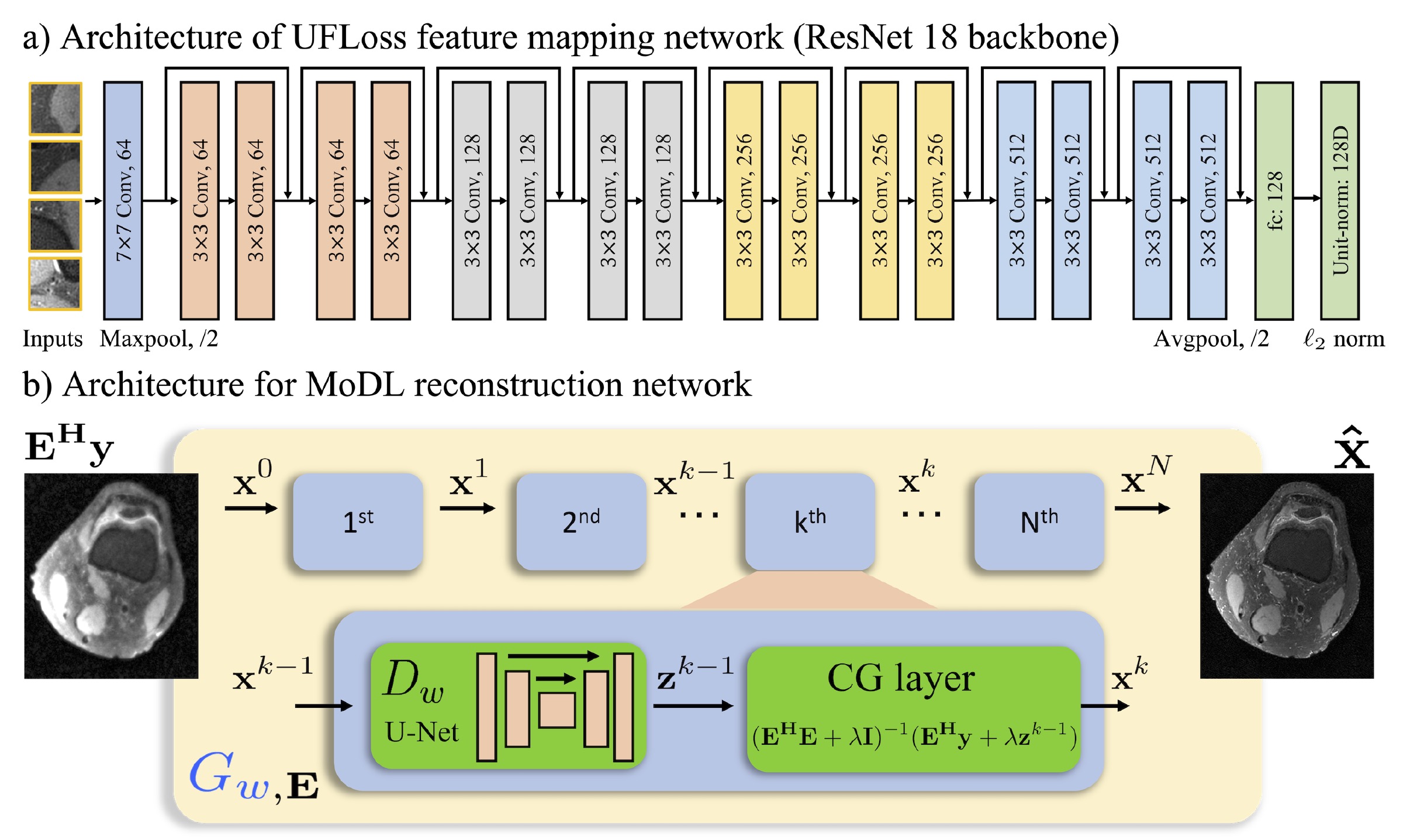}
\end{center}
\caption{a) The UFLoss feature mapping network is based on a ResNet 18 network structure \cite{he2016deep} and followed by an $\ell_2$ normalization layer to map the input patches to the 128D unit-norm feature space. b) Architecture of the MoDL \cite{aggarwal2018modl} reconstruction network. A data consistency Conjugate Gradient Descent (CG) module is inserted after a CNN-based denoiser $D_w$. $D_w$ follows the structure of U-Net \cite{ronneberger2015u} with two input channels that represent the real and imaginary parts of the complex-valued image data.}
\label{fig:NW}
\end{figure}

For the unrolled reconstruction network architecture, we used the structure from the MoDL paper \cite{aggarwal2018modl}, where a CG block was inserted after a CNN-based denoiser, and unrolled with a fixed number of iterations. In this work, we used 5 unrolls and 6 CG steps. As shown in Figure \ref{fig:NW}b), a U-Net \cite{ronneberger2015u} architecture was adopted for the CNN-based denoiser $D_w$.

The training of MoDL was performed by minimizing the proposed loss function $L_{Recon}$ (Equation \ref{eq:L_recon_full}) over the training set for 50 epochs, with an empirical weighting parameter $\mu = 1.5$, and Adam \cite{kingma2014adam} optimizer with a  learning rate of $1\mathrm{e}{-4}$.

To compute the UFLoss, patches are extracted on a grid across the image with 5-pixel strides in both vertical and horizontal directions.  At each training step, both output and reference images are randomly shifted from 0 to 5 pixels in the vertical and horizontal directions to eliminate blocking artifacts. In this work, we chose the weighting parameter to balance the values of $L_{MSE-all}$ and $L_{UFLoss-all}$ so that they are on par after the training converges. During inference, a zero-filled reconstruction is passed through the MoDL reconstruction network. Note that training with UFLoss does not change the network architecture, so the inference time remains the same as MoDL with pure $\ell_2$ loss.

All the proposed algorithms were implemented using Pytorch 1.2 \cite{paszke2017automatic}, and were run on 12GB Nvidia Titan Xp graphics processing units (GPUs).

\subsection{Evaluation of the proposed UFLoss}
% Could consider combining figure 4 and 5 in one subsubsection, or splitting figure 5 in 2 sections, etc

% Methods for Figure 4  - perturbations
\subsubsection{UFLoss as valid loss function}

To evaluate whether UFLoss is also a valid loss function for comparing two images at the intensity level, we study how the UFLoss changes with different sizes of perturbations in  two representative types:
\begin{enumerate}
\item Additive white Gaussian noise. 

A perturbed image $\mathbf{x_{p}}$ is generated from the original image $\mathbf{x_{o}}$ by adding different levels of additive Gaussian noise $\mathbf{n_\sigma}$:
\begin{linenomath*}
\begin{equation}
    \mathbf{x_{p}} = (1-\beta)\mathbf{x_{o}}+\beta\mathbf{n_\sigma},
    \label{GN}
\end{equation}
\end{linenomath*}
where $\beta$ is the noise level parameter in the range of $0-10\%$, and noise $\mathbf{n_\sigma}$ follows normal distribution: $\mathbf{n_\sigma}\sim \mathcal{N}(0,1)$.  We study how $L_{UFLoss}(\mathbf{x_{o}},\mathbf{x_{p}})$ changes as $\beta$ increases.   

\item Image blurring. 

A perturbed low-resolution image $\mathbf{x_{p}}$ is generated by cropping and zero-padding the k-space of the original image $\mathbf{x_{o}}$. The k-space cropping rate $\mathbf{R}$ ranges from 1-4. $\mathbf{R}=4$ indicates that only 25\% of k-space samples in both horizontal and vertical dimensions are kept. A higher $\mathbf{R}$ corresponds to more blurring and a coarser resolution.  We study how $L_{UFloss}(\mathbf{x_{o}},\mathbf{x_{p}})$ varies with different $\mathbf{R}$'s.
\end{enumerate}

In addition, we evaluate whether, by minimizing the objective UFLoss between the original and perturbed images $L_{UFloss}(\mathbf{x_{o}},\mathbf{x_{p}})$, we are able to guide the perturbed version towards the original version without falling into local minima. The starting perturbed image $\mathbf{x_{p-0}}$ is generated by image blurring where $\mathbf{R}=4$.  We update it per gradient descent with respect to $L_{UFloss}(\mathbf{x_{o}},\mathbf{x_{p-k}})$ in an iterative fashion: 
\begin{linenomath*}
\begin{equation}
    \mathbf{x_{p-k+1}}=\mathbf{x_{p-k}} - \alpha\frac{\partial L_{UFLoss}(\mathbf{x_{o}},\mathbf{x_{p-k}})}{\partial \mathbf{x_{p-k}}},
\end{equation}
\end{linenomath*}
where $\mathbf{x_{p-k}}$ is the perturbed image after $\mathbf{k}$ steps of gradient descent.

% Methods for Figure 5  - patch retrieval and correlation maps
\subsubsection{Perceptual Similarity}

In order to better interpret and understand the perceptual features learned for the UFLoss, we performed a patch retrieval experiment to evaluate and show patch pairs with high and low UFLoss feature similarities. First, we constructed a feature database (memory bank) by running all training patches through the pre-trained UFLoss network. Then, given an input patch from the testing set, we passed it through the network and queried its neighbors from the training patches based on their distances (inner products) in the feature space. We picked and visualized patches of the highest feature inner products with the input patch and also counter-examples with relatively low inner products.

% \textcolor{red}{do we want some math here? or reference some equation from before if already mentioned}
% \begin{equation}
%     similarity = f(x_1) \cdot f(x_2) \ \ or \ \ d = (f(x_1) - f(x_2)) ^ 2
% \end{equation}

% Correlation map

To further evaluate the UFLoss sensitivity and perceptual similarity for different anatomies and contrasts, we constructed correlation maps by computing the feature correlation (inner product) between a {\it source patch} and all patches in different images and visualized them as heatmaps. This experiment helps us better understand how anatomy and structure similarities relate to UFLoss feature similarities. 

Specifically, we first extracted a {\it source patch} from a source image. Then, we computed the feature correlations between the source patch and all patches on a grid from 1) the same source image; 2) the target image with the same contrast but from a different subject; and 3) the target image with different contrast and also from a different subject. Patches closer to the source patch in the feature space correspond to higher inner products.  We evaluated this experiment on both PDFS and PD scans. For comparisons, we also conducted the same experiments for the SSIM feature, where we computed the SSIM score between the source patch and all patches from different images.

% Methods for Figure 6, 7, 8, etc - reconstruction
\subsubsection{Unrolled Reconstructions with UFLoss}

To quantitatively evaluate our proposed UFLoss on under-sampled MRI reconstruction, we implemented both PICS \cite{lustig2007sparse} and MoDL \cite{aggarwal2018modl}.
In the unrolled reconstruction experiments, MoDL with our proposed UFLoss was compared with PICS and with MoDL using only per-pixel $\ell_2$ loss. The PICS method was implemented using the BART Toolbox \cite{uecker2015berkeley} with wavelets as the sparse transform. In order to further demonstrate the performance of our UFLoss, MoDL with $\ell_2$ + perceptual VGG loss \cite{yeh2016semantic} was also included in our comparisons.

For all the experiments, reconstruction performance was evaluated using different quantitative metrics, which reflect different aspects of image quality. The normalized root mean squared error (NRMSE) was used to measure the overall pixel-wise errors. SSIM \cite{wang2004image} was used to assess the local image similarity with respect to the fully sampled reference. At the same time, we also computed our proposed UFLoss between the reconstructed images and the fully sampled references.

\section{Results}
% \subsection{Interpretations of UFLoss} Any interpretrations go in discussion

% Add a section to demonstrate UFLoss can be indeed used as a loss function - could also go in discussion

\subsection{UFLoss as a valid loss function}

\begin{figure}[!ht]
\begin{center}
\includegraphics[width=14cm]{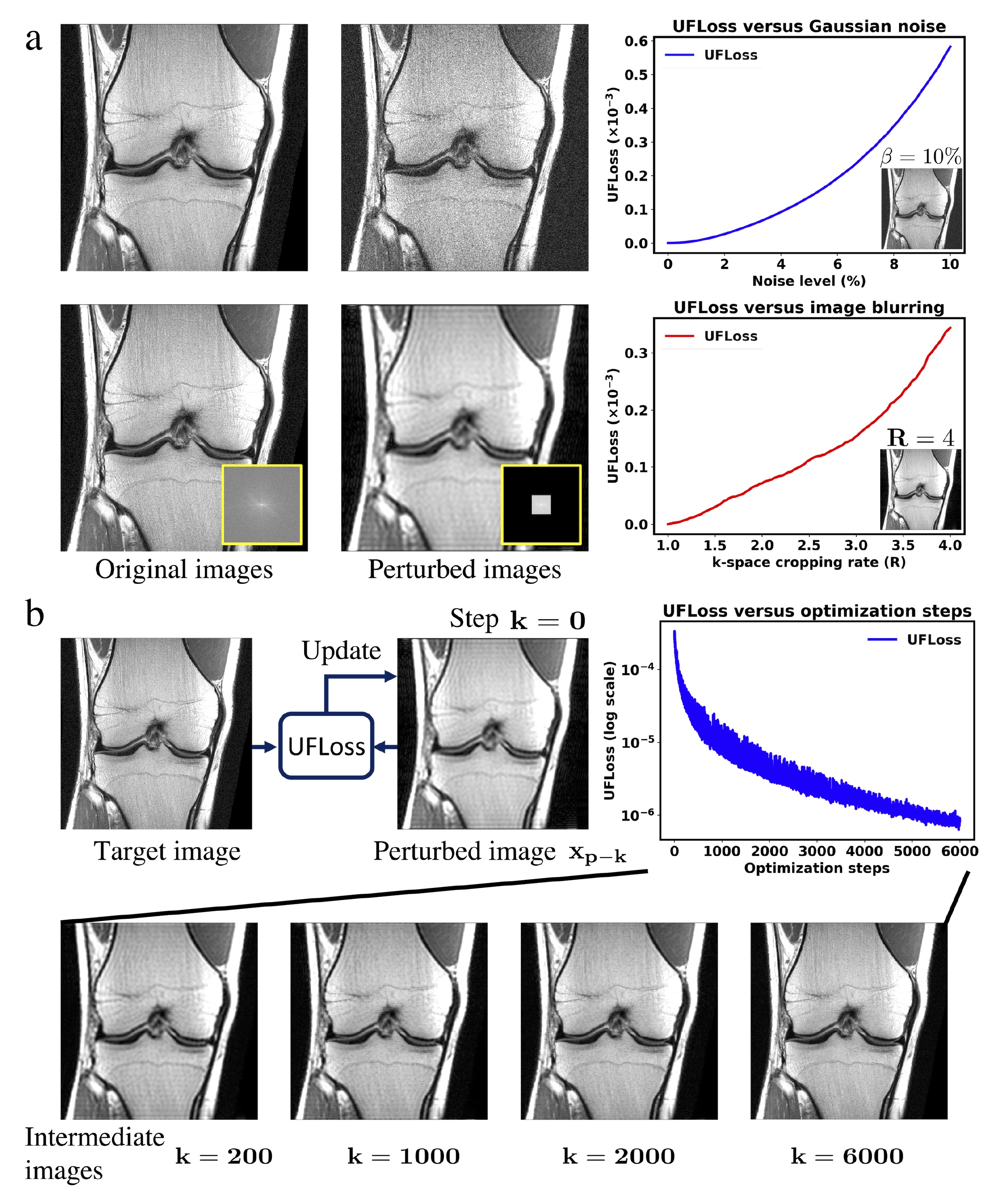}
\end{center}
\caption{a) Evaluation of UFLoss with different levels of perturbations. \textbf{Upper}: additional Gaussian noise, \textbf{Lower}: image blurring through k-space cropping. UFLoss evolution curves indicate that UFLoss increases in a convex way with respect to more Gaussian noise and increases in a near-convex way with respect to more blurring. b) Evaluation of UFLoss in guiding a blurred image $\mathbf{x_{p-0}}$ to the target high resolution image. Gradient descent is performed on $\mathbf{x_{p-k}}$ to reduce the UFLoss with respect to the target image in an iterative way.  Intermediate images show that UFLoss is able to gradually guide the blurred image to the target without falling into any local minimum.)}
\label{fig:valid_loss}
\end{figure}

% This can go more in discussion. Instead here we can just report objective findings from the figure.
Figure \ref{fig:valid_loss} indicates that our proposed UFLoss could be used as a valid loss function by itself. As shown in Figure \ref{fig:valid_loss}a), UFLoss between the perturbed and original clean images increases in a convex way with respect to more Gaussian noise and increases in a near-convex way with respect to more blurring. Even though the UFLoss feature mapping network is not specifically trained for any such perturbations, it learns low-level perceptual similarities between images, where a larger intensity perturbation corresponds to a larger UFLoss. On the other hand, Figure \ref{fig:valid_loss}b) indicates that by minimizing the UFLoss between the perturbed and target images, we are able to successfully restore the blurred image towards the clean one without falling into any local minimum.  Intermediate deblurred image samples are shown in the figure along with the UFLoss evolution curve.

\subsection{Perceptual Similarity}

\begin{figure}[!ht]
\begin{center}
\includegraphics[width=10cm]{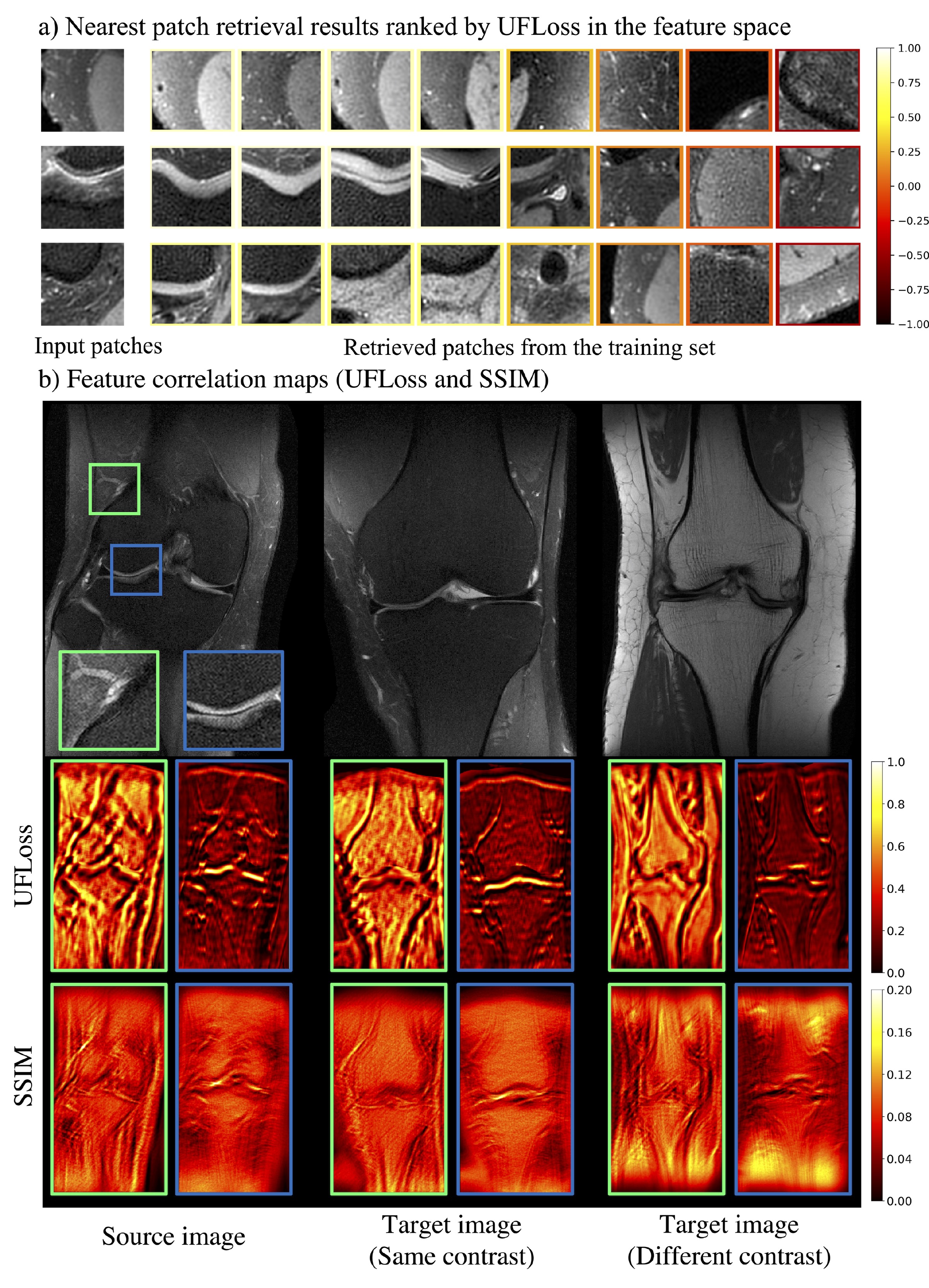}
\end{center}
\caption{Results after training the UFLoss feature mapping. a) Feature clustering results using UFLoss feature mapping where, given an input patch, neighbor patches from the training set can be queried based on their feature space distance. The left four patches are the closest neighbors with the input patch and have the highest inner products. At the same time, we also show four counterexamples with relatively low inner products with the input patch.
The feature space inner products between the input patch and the retrieved patches are shown as different colors of the borders.  The color bar on the right indicates that a brighter border corresponds to a higher correlation while a darker border corresponds to a lower correlation.  b) Feature correlations between different patches. The heat maps under a certain image show the feature correlations (feature space inner products for UFLoss) between all the patches from the image and the reference patches from the source image (first column). The heat maps with green/blue borders correspond to different source patches whose borders have the same colors. The correlation results for PDFS contrast using UFLoss and SSIM features are shown in the top and bottom rows, respectively.}
\label{fig:fl}
\end{figure}

% In order to demonstrate the impact of UFLoss for deep learning reconstruction, 
Figure \ref{fig:fl}a) shows the feature similarity results using the UFLoss feature. The feature space inner products between the input patch and the retrieved patches are shown as different colors of the borders. As seen in the figure, patches with similar perceptual structures (e.g., edges, bone structures) are mapped closer to each other in the feature space.

Figure \ref{fig:fl}b) (PDFS) and Supporting Figure S1 (PD) show the feature correlation maps (UFLoss and SSIM) between different patches.
Two {\it source patches}, indicated with green and blue edges, were chosen from each source image in the left column. The heatmaps under to each image, with corresponding green and blue edges, show the corresponding maps for each source patch from the source image. For the UFLoss results, we only show the positive inner products for visualization purposes, while in principle, the inner products range from -1 to 1. As shown in the UFLoss feature correlation maps, patches containing meniscus from both the same contrast and different contrast show high correlations with the input patch of the meniscus (blue border) while, on the other hand, patches from other anatomy show low correlation with it. These UFLoss feature correlation maps indicate that our unsupervised feature mapping is able to capture the perceptual structure similarities across different subjects and across different contrasts. In contrast, SSIM feature correlation maps do not successfully capture perceptual similarities across anatomies and contrasts (e.g., meniscus). More specifically, as shown in supporting figure S2, patch with the highest UFLoss feature correlation (top) shows very similar anatomical textures of the meniscus compared to the source patch. At the same time, because SSIM focuses more on the local signal statistics instead of high-level perceptual similarity, the patch with the highest SSIM (bottom) has totally different textures from a different anatomical region.

\subsection{Unrolled reconstructions with UFLoss}
\begin{figure}[!htb]
\begin{center}
\includegraphics[width=14cm]{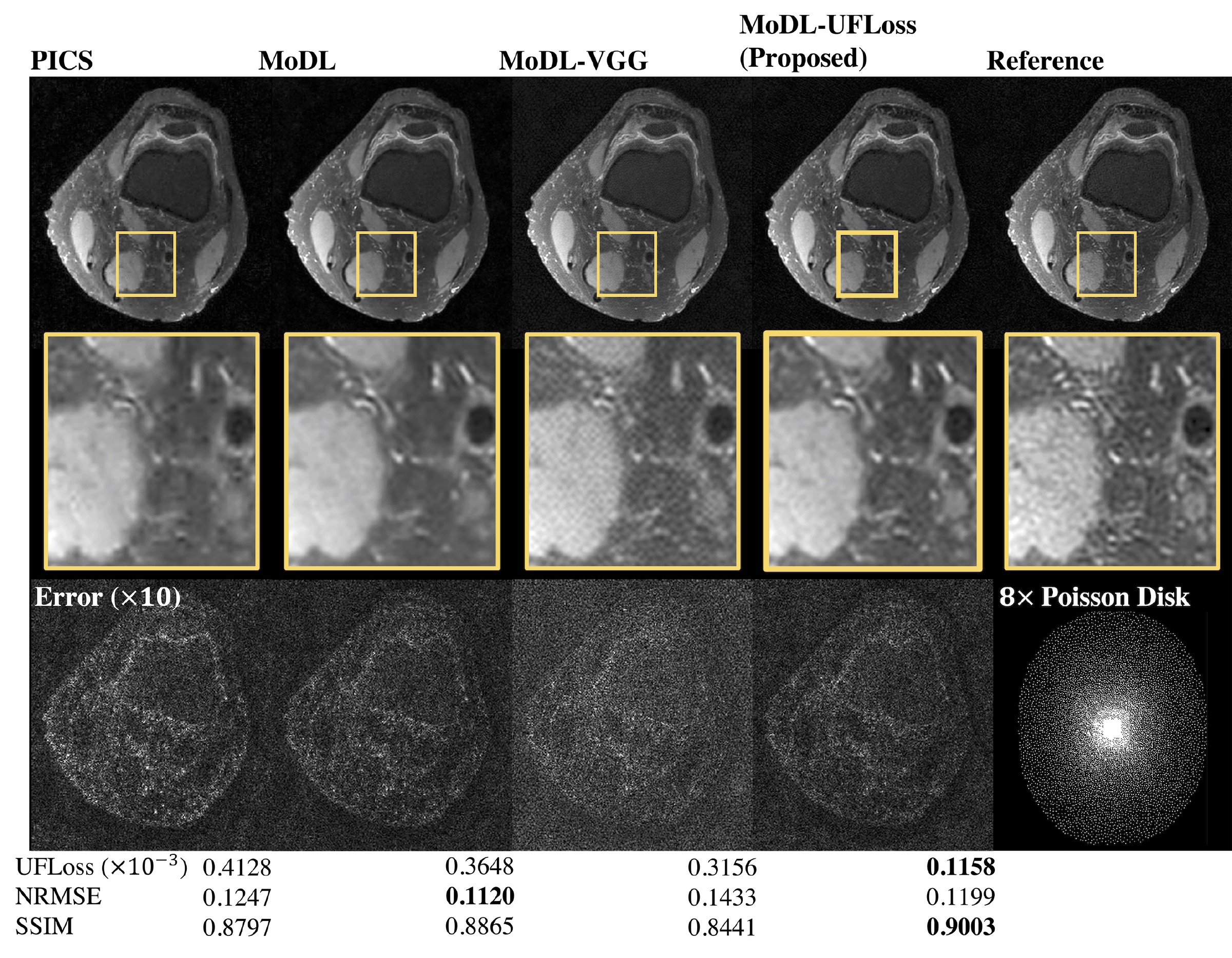}
\end{center}
\caption{Representative 3D knee reconstruction results from different methods. A fully-sampled scan is retrospectively under-sampled with a Poisson under-sampling mask by a factor of 8.  From left to right are reconstructions by: PICS, MoDL with $\ell_2$ loss, MoDL with $\ell_2$+perceptual VGG loss, and MoDL with $\ell_2$+our proposed UFLoss. NRMSE, SSIM, and UFLoss for each method are computed with respect to the fully sampled reference and shown under the image for reference. As shown in the zoomed images and error maps, our proposed MoDL with UFLoss showed sharper edges and more detailed structures with high perceptual similarity compared to the reference image. }
\label{fig:3DRC}
\end{figure}

Figure \ref{fig:3DRC} shows reconstruction comparisons between different methods (PICS, MoDL, MoDL with VGG, MoDL with UFLoss) for a representative 3D knee scan with under-sampling rate of $R=8$. Quantitative metrics (NRMSE, SSIM) are shown under the images. As indicated in the zoomed images and error maps, MoDL with UFLoss shows finer structural details, sharper edges, and higher perceptual agreement with the fully-sampled reference images compared to the other reconstruction methods. Without our UFLoss, pure $\ell_2$ loss at this under-sampling rate leads to blurring and perceptual quality degradation. MoDL with the VGG perceptual loss \cite{johnson2016perceptual} shows higher perceptual quality compared with MoDL, but generates unintended checkerboard structured artifacts, which is consistent with findings in \cite{sugawara2018super,odena2016deconvolution}.

\begin{figure}[!htb]
\begin{center}
\includegraphics[width=14cm]{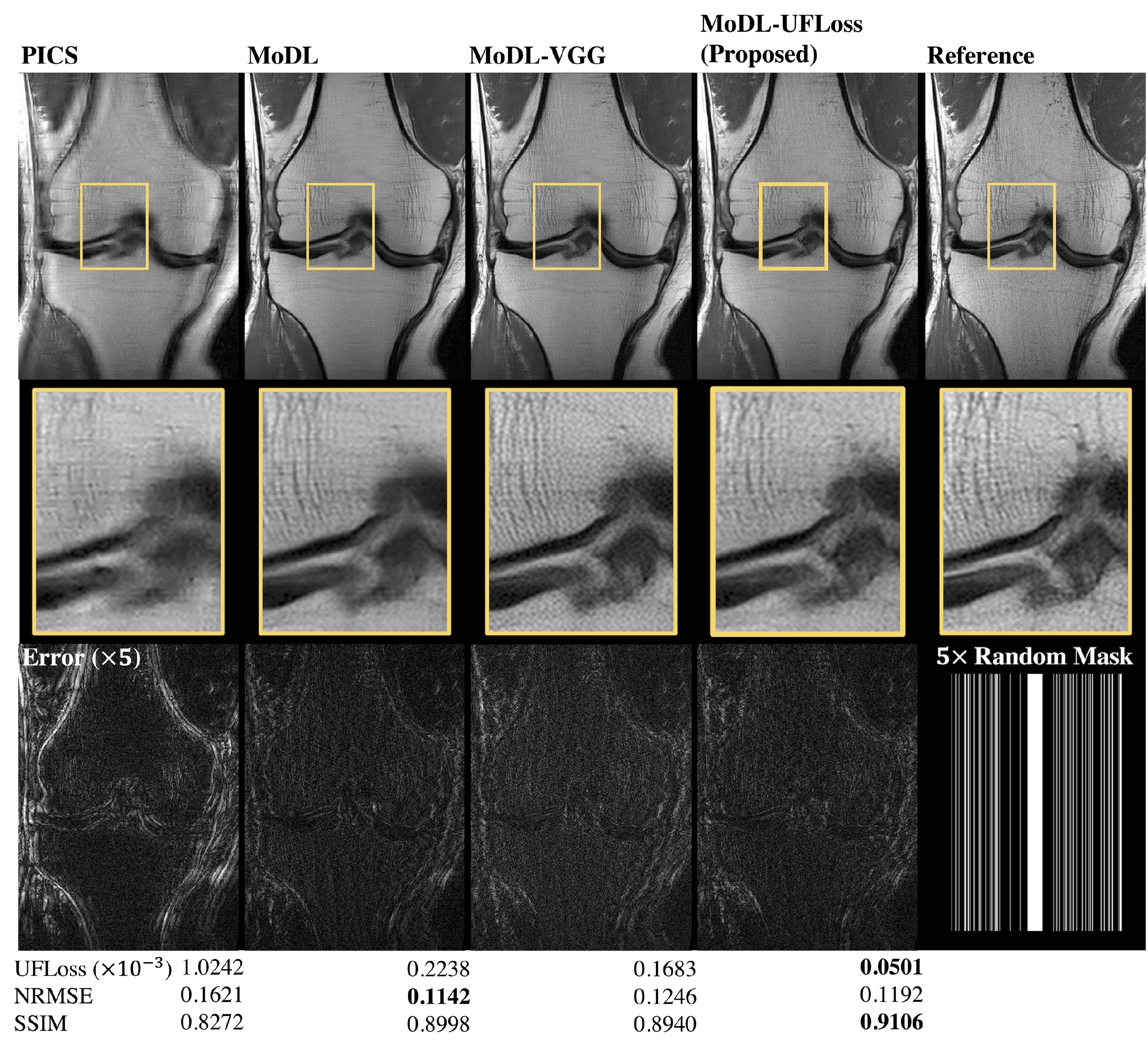}
\end{center}
\caption{Representative examples of 2D PD knee reconstruction results using different methods. A fully-sampled slice is retrospectively randomly under-sampled by a factor of 5. From left to right are reconstructions by PICS, MoDL with $\ell_2$ loss, MoDL with perceptual VGG loss, and MoDL with our proposed UFLoss. NRMSE, SSIM, and UFLoss for each method are shown below the figure for references. As shown in the zoom-in views and error maps, our proposed MoDL with UFLoss can provide more realistic and natural-looking textures, while MoDL with $\ell_2$ loss alone tends to blur out some high-frequency textures.}
\label{fig:2DRC_NOFS}
\end{figure}

Figure \ref{fig:2DRC_NOFS} shows the comparison of different reconstruction methods for a representative 2D PD slice from the fastMRI dataset  \cite{zbontar2018fastmri}. The retrospective 2D under-sampling rate is 5, where around 20\% of the k-space data is sampled. At this acceleration rate, PICS failed to effectively recover the fine bone structures, and MoDL with  $\ell_2$ loss alone also suffers blurring artifacts. In contrast, MoDL with UFLoss demonstrates more realistic reconstruction performance with more detailed texture everywhere, including the bone.

\begin{figure}[!htb]
\begin{center}
\includegraphics[width=14cm]{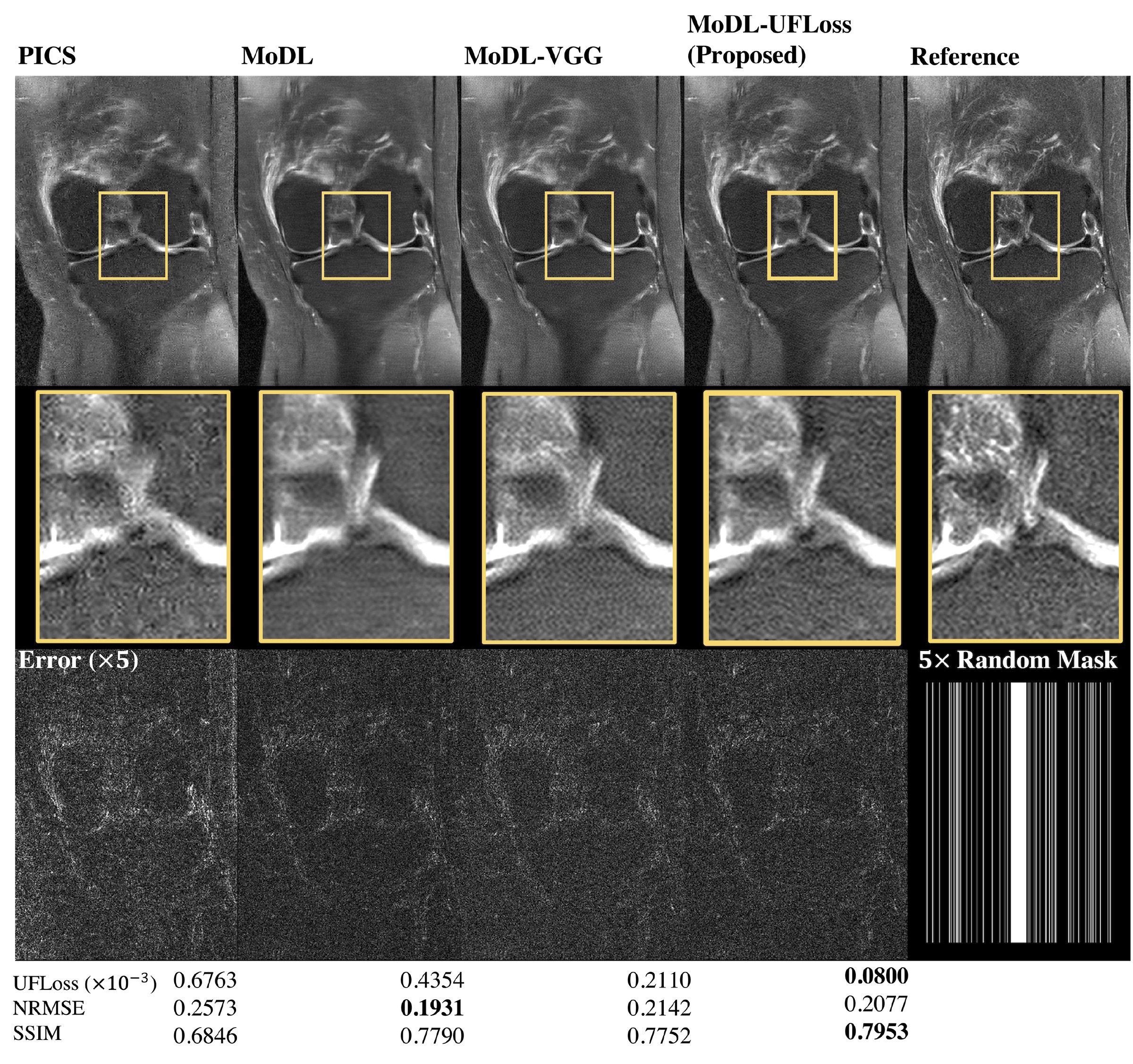}
\end{center}
\caption{Representative examples of 2D PDFS knee reconstruction results using different methods at under-sampling rate R=5. nRMSE, SSIM, and UFLoss for each method are shown in the figure.  Quantitative metrics indicate that MoDL with UFLoss has the highest SSIM and the lowest UFLoss, as well as the highest perceptual quality of the reconstructed image. Meanwhile, as shown in the zoom-in images and error maps, our proposed MoDL with UFLoss reconstruction looks more natural with a more faithful contrast than other methods.} 
\label{fig:2DRC_FS}
\end{figure}

Figure \ref{fig:2DRC_FS} shows the reconstruction comparisons for a representative 2D PDFS slice from the fastMRI dataset \cite{zbontar2018fastmri}. Quantitative comparisons are shown at the bottom of the figure.  Due to the suppression of the fat signal, the SNR of the data is relatively low, where high-frequency features can be mixed up with the noise. The zoomed-in views and the corresponding error maps indicate that PICS results in a high level of artifacts. Meanwhile, MoDL with $\ell_2$ loss alone misses fine detailed structures. Similar to the analysis above, MoDL with the VGG feature loss is capable of recovering subtle structures but generates unintended structured artifacts. In contrast, MoDL with UFLoss can effectively recover the detailed texture and have the most realistic reconstructions.

So far, for all of our experiments, we used a fixed UFLoss weighting factor ($\mu=1.5$) for Equation \ref{eq:L_recon_full}. Supporting Figure S3 shows two representative reconstruction results with different UFLoss weighting factors during the training. We can clearly see that neither pure $\ell_2$ loss nor pure UFLoss achieves the best image quality. By combining these two terms, our model is able to take advantage of both the per-pixel intensity information and patch-level perceptual similarities.

\begin{figure}[!htb]
\begin{center}
\includegraphics[width=16cm]{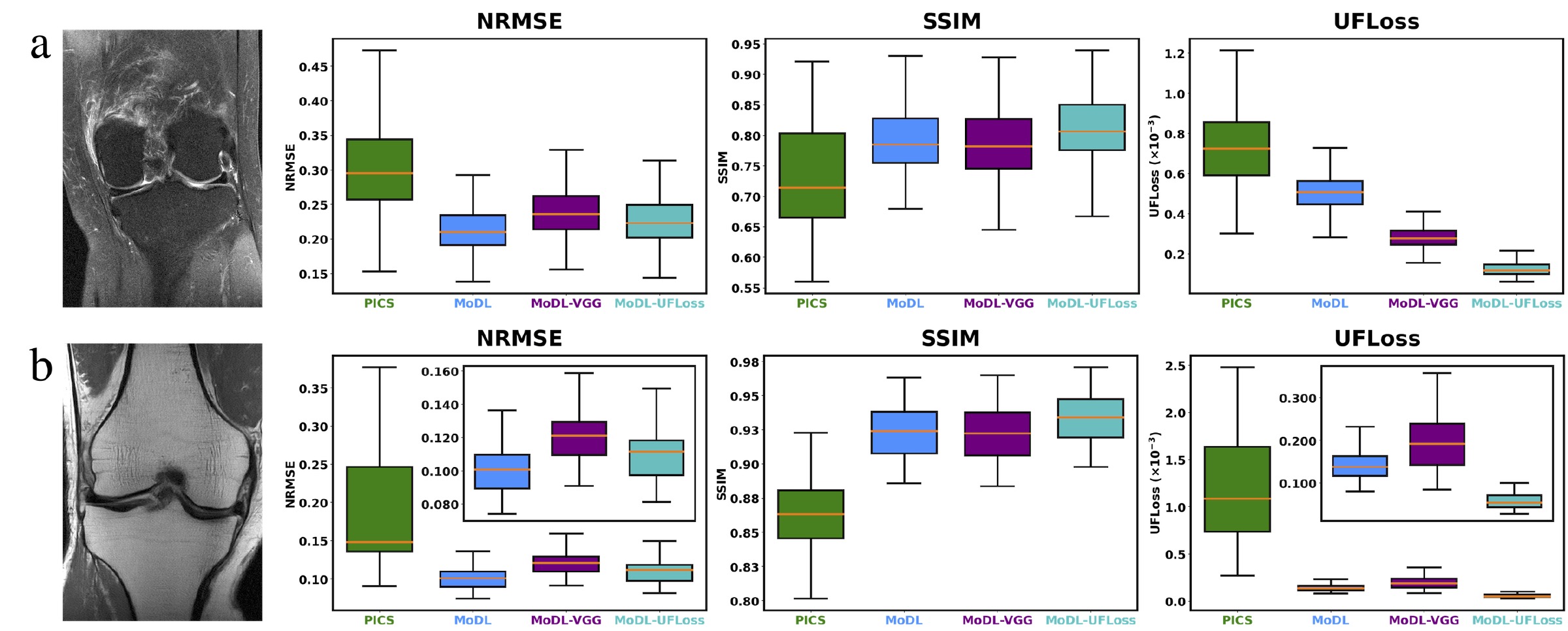}
\end{center}
\caption{Box plots for the metric comparisons. Two representative fully-sampled scans (10 PD and 10 PDFS) with 15 slices each are randomly under-sampled by a factor of 5 and reconstructed using PICS, MoDL, MoDL with perceptual VGG loss, and MoDL with UFLoss. NRMSE, SSIM, and UFLoss are calculated with respect to fully sampled reference images and shown in the plot. We use zoomed-in plots to show more clear comparisons for some sub-plots.}
\label{fig:Metrics}
\end{figure}

Figure \ref{fig:Metrics} shows the quantitative metric (NRMSE, SSIM, UFLoss) comparisons for the 2D unrolled reconstruction experiments. For both a) PD and b) PDFS experiments, ten representative testing scans with 15 slices each are used to calculate the quantitative metrics. As indicated in the figure, for both contrasts, MoDL with UFLoss outperforms both PICS and MoDL with $\ell_2$ loss in terms of SSIM and UFLoss and can achieve comparable performance in terms of NRMSE.

\section{Discussion}
In this work, we presented a novel patch-based perceptual loss function, which we call Unsupervised Feature Loss or UFLoss. UFLoss corresponds to the $\ell_2$ distance in a low dimensional feature space. Feature vectors are mapped from image patches through a pre-trained mapping network. The mapping network aims to maximally separate all the patches in the feature space, where similar patches become closer to each other, capturing high-level perceptual similarities.  As indicated in Figure \ref{fig:fl}, unlike $\ell_2$ distance, which focuses on the pixel-wise values, our proposed UFLoss agrees better with human visual judgment, where similar-looking patches have lower UFLoss in the feature space. By incorporating UFLoss into the training of DL-based reconstructions, we are able to recover finer textures, smaller features, and sharper edges with higher overall image quality compared to conventional per-pixel losses. 
% Unlike other common feature losses (e.g., perceptual VGG loss) that are trained with human-annotated semantic category labels, the training of the mapping network does not require any human annotation, which is difficult to define and obtain for medical images. 
% By using a ResNet \cite{he2016deep} based backbone, our proposed unsupervised feature mapping is able to capture the high-level perceptual similarity, which cannot be achieved by simple per-pixel losses. 
By leveraging a memory bank to store all the features, the training of our mapping network becomes feasible for a large dataset: The UFLoss network training required less than 500 MB GPU memory and was easily trained within two hours.

As we mentioned before, another important class of feature losses for DL-based reconstruction is adversarial loss or GAN loss \cite{goodfellow2014generative}. Adversarial losses have shown great success in capturing perceptual properties of ground-truth images and could be used to improve the reconstruction quality.  However, this loss is an instance-to-set loss, where a group of images can have similar low adversarial losses with respect to a certain image. Those losses have been shown to be less stable in reconstruction scenarios and can generate unintended hallucinations \cite{cohen2018distribution,muckley2021results,edupuganti2020uncertainty}. In comparison, our proposed patch-based UFLoss is an instance-to-instance loss, which is more constraining during training and likely more robust to those hallucinations.

% On the other hand, different from perceptual VGG loss, which are global losses, our proposed UFLoss works on image patches (e.g., $40\times40$ patches in a $256\times320$ image). Powered by overlapping patches, unlike other feature losses, UFLoss can be effectively used as a proper loss function by itself. The extracted features mainly focus on local textures and structures, which are more stable and can effectively reduce the variation induced by global feature losses.

In this study, UFLoss can be viewed as a separate module and be easily incorporated into other learning frameworks. The performance of UFLoss was demonstrated for accelerating 2D and 3D knee imaging by comparing the reconstruction results with respect to fully sampled references. The in-vivo results show that the addition of UFLoss during the network's training allows realistic texture recovery and improves overall image quality compared to a reconstruction network trained without UFLoss.

\begin{figure}[!htb]
\begin{center}
\includegraphics[width=15cm]{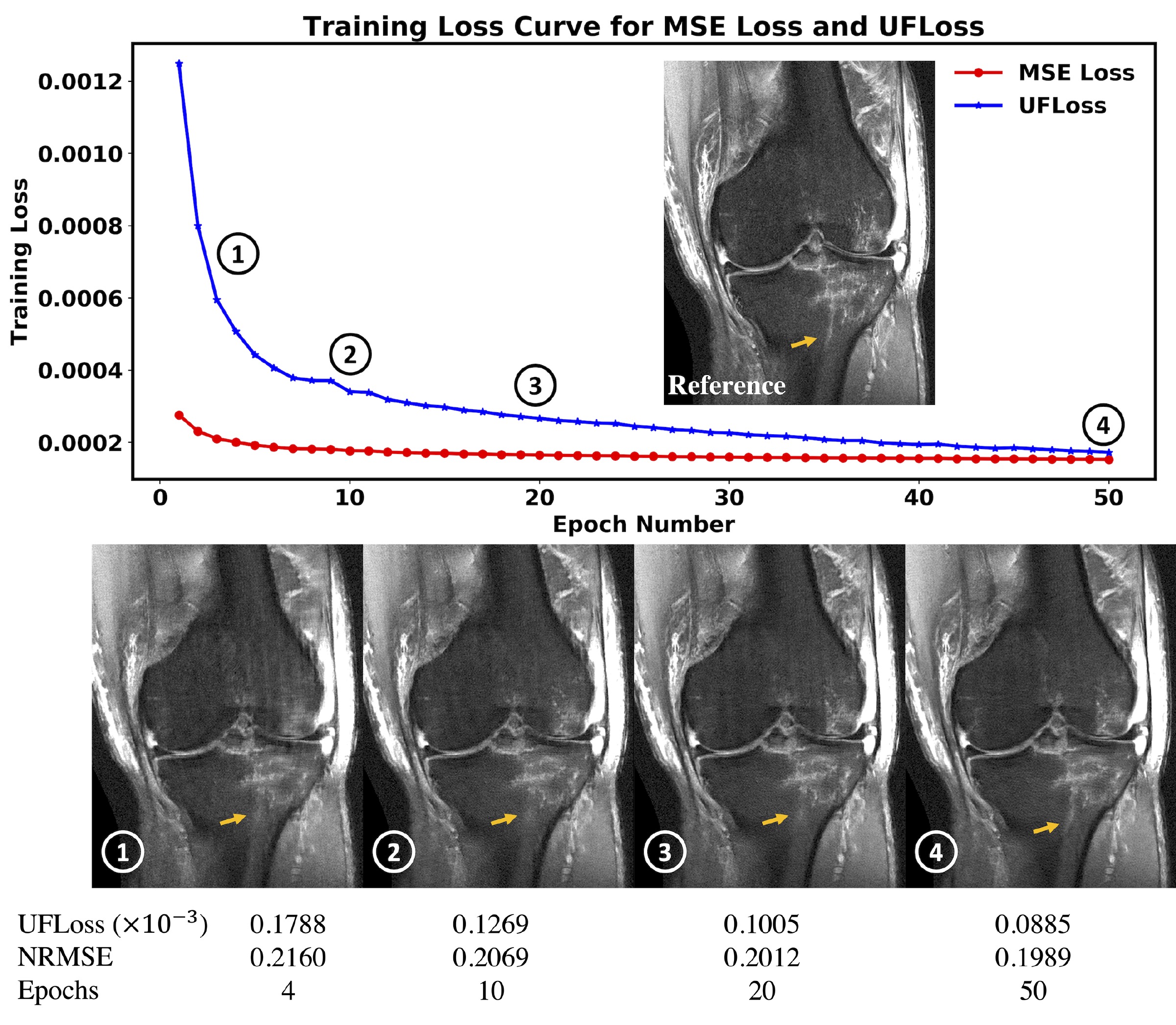}
\end{center}
\caption{Training loss curves for the $l_2$ MSE loss and our proposed UFLoss. A 2D fully-sampled slice is randomly under-sampled by a factor of 5 and reconstructed at different training epochs. NRMSE and UFLoss are shown as quantitative metrics under each reconstructed image. Yellow arrows point at the same representative textures at different reconstructions.}
\label{fig:lc}
\end{figure}

Another interesting finding of the UFLoss comes from how the training losses evolve, as shown in Figure \ref{fig:lc}.  The total loss consists of two different components, the per-pixel $\ell_2$ MSELoss and our proposed UFLoss, which are shown in the top sub-figure as red and blue curves, respectively. The bottom sub-figure shows the testing reconstruction results at different epochs. As indicated from the curve, the MSELoss remains almost constant after ten epochs, while our proposed UFLoss still decreases continuously.  Inspecting the reconstructed images at different training epochs, we can see that the image quality continues to improve with the further reduction of the UFLoss.  At the same time, the quantitative metrics indicate that those reconstructed images have very similar NRMSE compared with the fully-sampled reference but a much more significant difference in their UFLoss values. A low UFLoss value corresponds to better image quality. These results indicate that using the $\ell_2$ MSE loss alone is not optimal. Therefore, the UFLoss can be potentially used as a better perceptual comparison criterion and help further improve the reconstruction quality. 

One limitation of this study is that the training of DL-based reconstructions with UFLoss is time-consuming and memory-inefficient due to the extraction and feed-forwarding of a large number of patches within a single step. This can be potentially improved by using fully-convolutional image-scale networks and GPU parallel computing. On the other hand, we haven't thoroughly investigated the sensitivity of different hyperparameters (e.g., patch size, temperature parameter, UFLoss network depth) to the training and final reconstructions. Supporting Figure S3 demonstrates how UFLoss weighting parameter contributes to the reconstruction results. A more thorough parameter search and analysis will be explored in the future.

\section{Conclusion}

In summary, a novel patch-based feature loss, {\it Unsupervised Feature Loss} or \textit{UFLoss}, is proposed, and it can be easily incorporated into the training of any existing DL-based reconstruction frameworks without any modification to the model architecture. UFLoss is based on an unsupervised pre-trained feature mapping network without any external supervision. With the addition of our proposed UFLoss, we are able to reconstruct high fidelity images with sharper edges, more faithful contrasts, and better image quality overall.

% \newpage

\section{Acknowledgment}

The authors thank Anja Brau, Sangtae Ahn, Graeme C McKinnon, Marc Lebel, Xucheng Zhu, Gopal Nataraj and Efrat Shimron for their helpful suggestions, and Efrat Shimron for her help with the paper editing. 
% We also acknowledge support from NIH grants R01EB026136, R01HL136965, R01EB009690 and GE Healthcare.

\section{Data Availability Statement}
In the spirit of reproducible research, our source code can be found at  \url{https://github.com/mikgroup/UFLoss} to reproduce most of the results in this paper.

% \section{Conflict of Interest}
% Our group receives research support from GE Healthcare. Uri Wollner and Rafi Brada are employees of GE Global Research.

% \section{ORCID}
% \textit{Ke Wang} \url{https://orcid.org/0000-0001-5951-1727}

{\renewcommand{\thefigure}{S1}
\begin{figure}[!ht]
\begin{center}
\includegraphics[width=13cm]{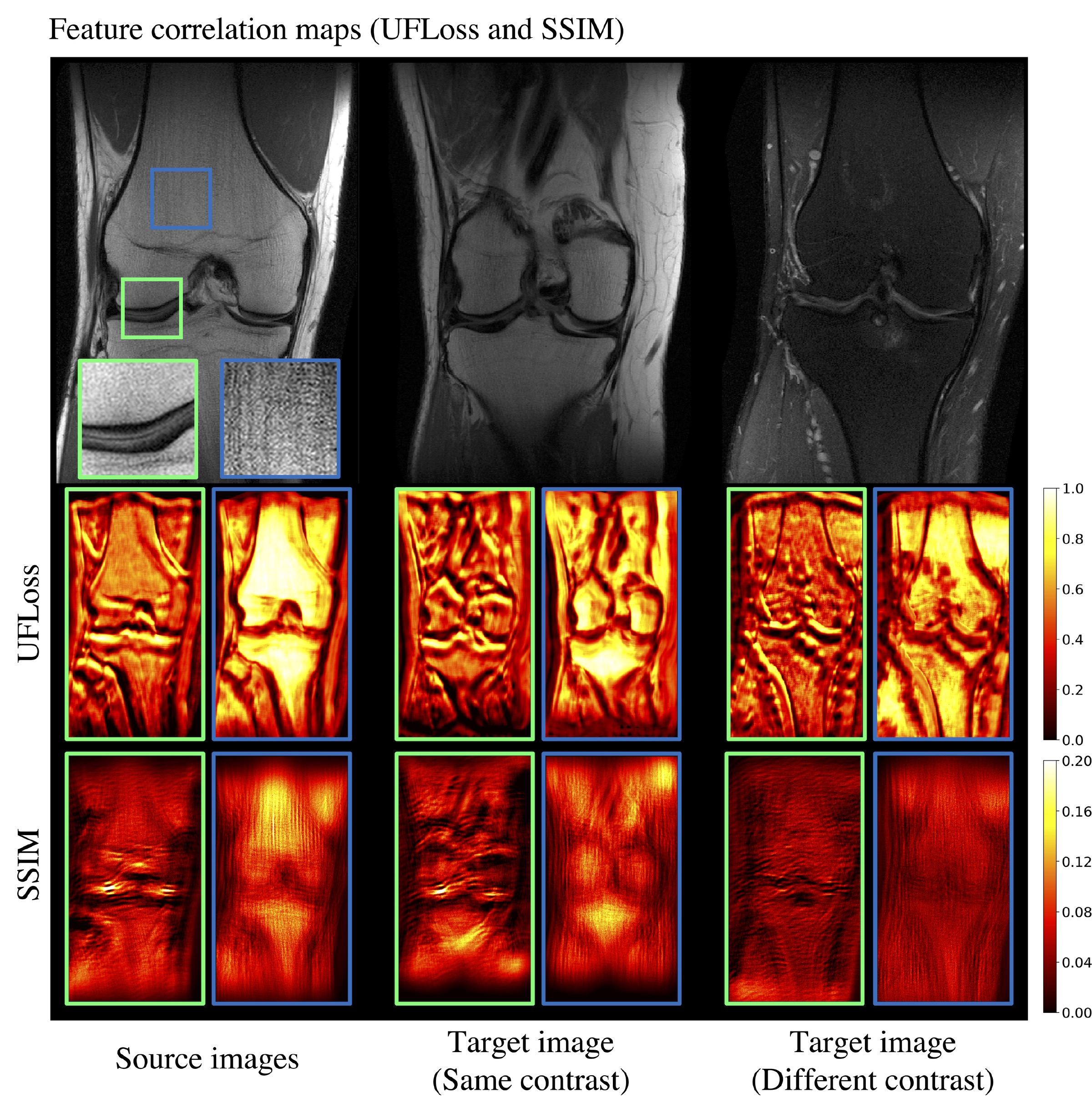}
\end{center}
\caption{Feature correlations between different patches. The heat maps under a certain image show the feature correlations between all the patches from the image and the source patches from the source image (first column). The heat maps with green/blue borders correspond to different source patches whose borders have the same colors. The correlation results for PD contrasts using UFLoss and SSIM features are shown in the top and bottom rows, respectively.}
\label{fig:fl_pdfs}
\end{figure}}

{\renewcommand{\thefigure}{S2}
\begin{figure}[!ht]
\begin{center}
\includegraphics[width=13cm]{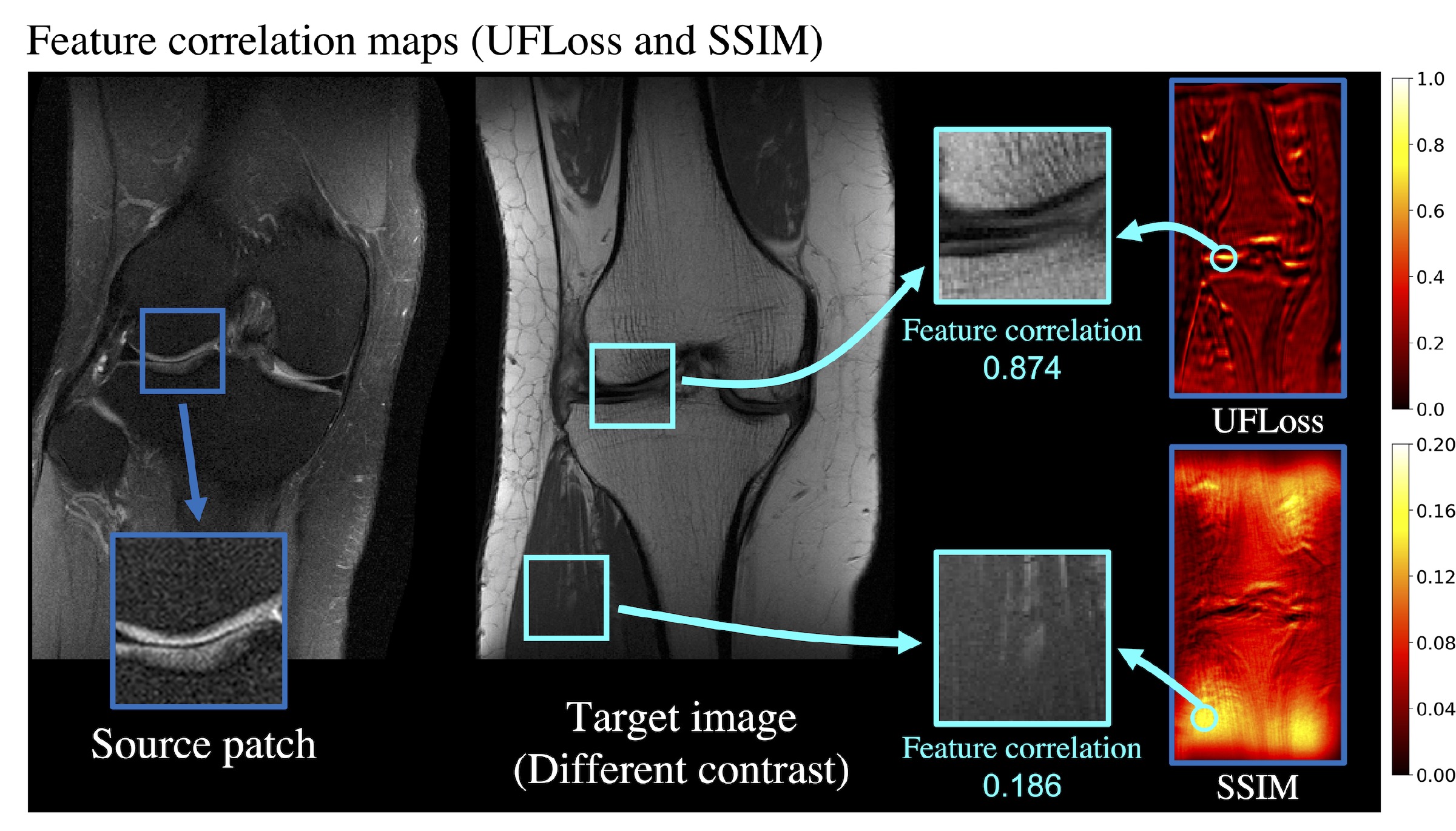}
\end{center}
\caption{Feature correlations between different patches. The heat maps alongside the PD image show the feature correlation values between all the patches from the PD image and the source patch from the PDFS image (first column). The correlation results using UFLoss and SSIM features are shown on the right. Patches with the highest UFLoss and SSIM feature correlations in the PD image are visualized as zoomed-in patches with light blue borders. Feature correlation value are shown under each patch.}
\label{fig:fl_ssim_r}
\end{figure}}

{\renewcommand{\thefigure}{S3}
\begin{figure}[!htb]
\begin{center}
\includegraphics[width=16cm]{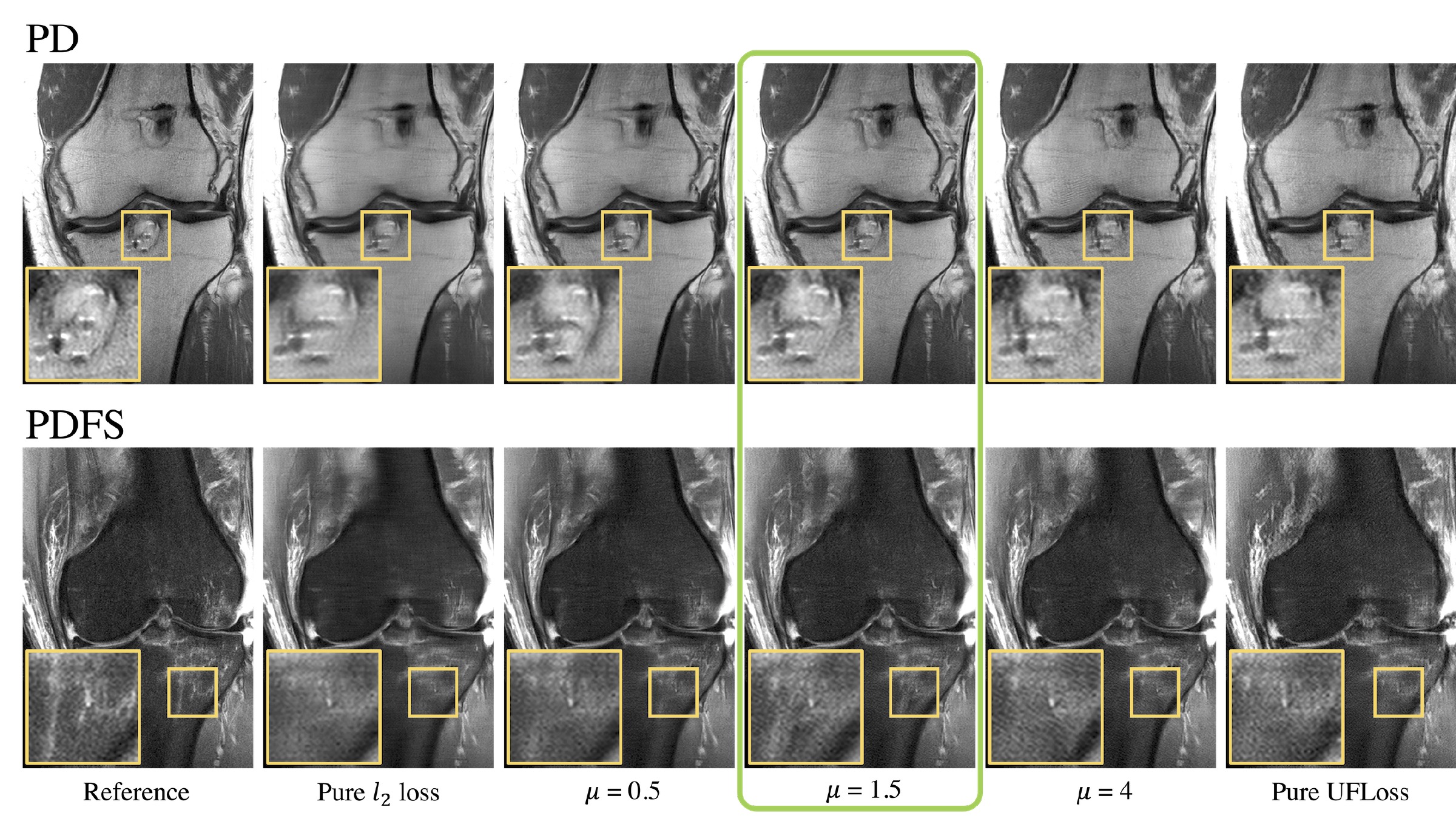}
\end{center}
\caption{Representative examples of 2D PD and 2D PDFS knee reconstruction with different UFLoss weighting factors during the training. Fully-sampled slices are retrospectively randomly under-sampled by a factor of 5, and reconstructed using MoDL with different weights of UFLoss. Pure $\ell_2$ loss, combined $\ell_2$ and UFLoss with $\mu$=0.5,1.5,4, and pure UFLoss are included for evaluations. Zoomed-in details are shown along with each image.} 
\label{fig:Weightings}
\end{figure}}

\clearpage
\bibliographystyle{ama}
\bibliography{sample}

\clearpage
\newpage
\section{List of Supporting Information Figures}
\textbf{S1} Feature correlations between different patches. The heat maps under a certain image show the feature correlations between all the patches from the image and the source patches from the source image (first column). The heat maps with green/blue borders correspond to different source patches whose borders have the same colors. The correlation results for PD contrasts using UFLoss and SSIM features are shown in the top and bottom rows, respectively.

\vspace{\baselineskip}
\noindent\textbf{S2} Feature correlations between different patches. The heat maps alongside the PD image show the feature correlation values between all the patches from the PD image and the source patch from the PDFS image (first column). The correlation results using UFLoss and SSIM features are shown on the right. Patches with the highest UFLoss and SSIM feature correlations in the PD image are visualized as zoomed-in patches with light blue borders. Feature correlation value are shown under each patch.

\vspace{\baselineskip}
\noindent\textbf{S3} Representative examples of 2D PD and 2D PDFS knee reconstruction with different UFLoss weighting factors during the training. Fully-sampled slices are retrospectively randomly under-sampled by a factor of 5, and reconstructed using MoDL with different weights of UFLoss. Pure $\ell_2$ loss, combined $\ell_2$ and UFLoss with $\mu$=0.5,1.5,4, and pure UFLoss are included for evaluations. Zoomed-in details are shown along with each image.

\end{document}